# Evidence for Anomalies in Muon-Induced Neutron Emissions from Pb

**W. H. Trzaska,** [a, b, *] **A. Barzilov,** [c] **T. Enqvist,** [a] **K. Jedrzejczak,** [d] **M. Kasztelan,** [e] **P. Kuusiniemi,** [a] **K. K. Loo,** [a] **J. Orzechowski,** [e] **M. Słupecki,** [f] **J. Szabelski,** [d] **and T. E. Ward** [g, h]

[a] *Department of Physics, University of Jyväskylä, Finland*
[b] *Helsinki Institute of Physics (HIP), University of Helsinki, Finland*
[c] *University of Nevada Las Vegas, Department of Mechanical Engineering, USA*
[d] *Stefan Batory Academy of Applied Sciences, Skierniewice, Poland*
[e] *Łódź, Poland; formerly at National Centre for Nuclear Research (NCBJ), Poland*
[f] *CERN, Switzerland*
[g] *Office of Nuclear Energy, DOE, USA*
[h] *TechSource, Santa Fe, NM, USA*

Abstract

This paper examines neutron multiplicity spectra emitted from massive targets at depths of 3, 40, 210, 583, 1166, and 4000 m.w.e. The measurements, carried out between 2001 and 2024, used three experimental setups with either 14 or 60 He-3 neutron detectors and lead (Pb) targets weighing 306, 565, or 1134 kg. The total acquisition time exceeded six years. When available, the spectra obtained were compared with Monte Carlo simulations. Our data indicate potential anomalies that may limit the accuracy of modelling muon-induced neutron multiplicity spectra using a single power-law function. Even at shallow depths, the single-power-law approach fails to account for a small but statistically significant excess of events at the highest multiplicities. This excess resembles a second power-law component. Since the anomaly varies only slightly with depth, it is unlikely to be directly linked to the muon flux. Our highest-quality data, acquired at 583 m.w.e., suggest a possible structure similar to the emission of approximately 74, 106, 143, and 214 neutrons from the target. We propose new underground measurements using low-cost, large-area, position-sensitive neutron arrays surrounding multi-ton Pb targets to verify and investigate these suspected anomalies and to determine their origin.

[*] Corresponding author. *E-mail:* wladyslaw.h.trzaska@jyu.fi

# 1 Introduction

Energetic cosmic rays reaching Earth interact with the upper atmosphere, generating Extensive Air Showers (EAS) of cascading subatomic particles and ionised nuclei that spread in the direction of the initial projectile. Most constituents of EAS are absorbed in the air or in the first 80 meters of water equivalent (m.w.e.) of the soil; however, neutrinos and high-energy muons penetrate deep underground. In most instances, neutrino interactions can be disregarded, while muons continue producing secondary showers in the rock and other materials they encounter along their path. They are, in addition to radioactive decay, the primary source of neutron events detected in deep locations. Because many cutting-edge experiments must account for this background, several studies, primarily based on simulations, have focused on this topic. For a comprehensive evaluation of muon-induced neutron production in lead, see, for instance [1].

The production of neutrons due to muonic bremsstrahlung is the dominant reaction in muon-matter interactions, resulting in neutron multiplicities ranging from $M$=1 to ~100 in lead. Conversely, photonuclear hadronic interactions induce nucleon spallation reactions in lead, yielding substantial neutron multiplicities, ranging from $M$=1 to ~200. The intensity of muon-induced hadronic reactions at 80 GeV and 80 TeV is approximately 1% and 2%, respectively, of the total muon interactions, correlating with the muon intensity at depth [2,3]. The computed neutron spectra exhibit long tails that extend to very high multiplicities [4]. With the decreased muon flux at deeper locations, the simulated neutron multiplicity spectra diminish in intensity but maintain their spectral shape. While most background studies rely on Monte Carlo (MC) simulations, their ability to predict subtle effects remains limited, as exemplified by recent GEANT4-based MC studies [5,6,7] that show discrepancies between measured and calculated values of muon-induced neutron yields.

The main limitation of current studies on muon-induced neutron yields in various materials, including lead, is the lack of experimental data. Although the general features of underground neutron multiplicity spectra are well understood, there is a shortage of long-exposure, high-precision data with sufficient sensitivity to examine the presence and characteristics of potential anomalies, especially at the high-multiplicity end of the spectra. New large-scale experimental datasets would offer vital reference points for simulations and aid in verifying interaction cross-sections. This paper partly addresses this issue.

Particularly interesting and relevant are potential anomalies, which could be an indirect sign of Dark Matter (DM) interactions or other exotic phenomena. We have thus analysed and searched for anomalies in the results of six measurements carried out over the past two decades, involving three experimental setups at different locations and depths. Data at 3 m.w.e. were obtained from the Khloplin Radium Institute in St. Petersburg, Russia. Data at 40 m.w.e. come from the underground laboratory of the Cosmic Ray Laboratory at the National Centre for Nuclear Research (NCBJ) in Łódź, Poland. Data at 210, 583, 1166, and 4000 m.w.e. were gathered in the Pyhäsalmi mine in Finland. The total data collection time for the analysed experiments exceeds six years. Partial results from these measurements were presented at conferences and appeared in the proceedings [8,9,10,11], but were not published earlier because, due to the low statistics (2-4 sigma), none of the individual experiments were conclusive. However, when evaluated collectively, they form a consistent case for the possible existence of anomalies in the muon-induced neutron multiplicity spectra emitted from lead at depths down to 4000 m.w.e.

The most prevalent theoretical models predict that DM is a Weakly Interacting Massive Particle (WIMP) with a mass in the GeV range. If such particles exist, they must scatter on normal matter. In practice, all terrestrial DM searches seek the tiny signals expected from particles recoiling after a WIMP collision, a direct interaction. So far, no such events have been identified [12,13]. However, if WIMPs exist, they should also decay or annihilate, albeit with a small cross-section. Many theorists expect WIMPs to be Majorana particles [14]. In that case, they would require another WIMP to annihilate. Nonetheless, it is prudent to consider and verify alternative scenarios, such as dark matter annihilation with baryonic matter. If this occurs within a massive target, the resultant high-energy gamma rays, energetic leptons, and particularly neutrons could provide a detectable signature of such a phenomenon. Thus, one should investigate potential anomalies in the neutron multiplicity spectra emitted from massive lead (Pb) targets situated at various depths, as was first proposed in 2002 [15]. Pb is preferred due to its large atomic number, high density, and availability. Moreover, while Pb is an effective absorber of gamma rays and charged particles, it is nearly transparent to neutrons, having a low absorption cross section for thermal neutrons (152 mb; resulting in an absorption mean-free path of ~2 m) and a high elastic scattering cross section for 1-MeV neutrons (11 b) [16].

# 2 Low-statistics Techniques

As the flux of cosmic-ray-induced muons decreases rapidly with depth, obtaining statistically significant spectra requires large targets, prolonged exposures, and extensive detector arrays. Despite this, the power-law dependence of occurrence frequency on neutron multiplicity presents a challenge for data analysis, particularly when comparing spectra from different measurements. Below, we discuss mitigation techniques that have assisted us in addressing these issues.

## 2.1 Spectra Integration

Using integrated spectra instead of original spectra is a common method to manage low statistics. The integrated spectrum is obtained by adding to each channel all events in the channels above. In other words, the number of events in the integrated neutron spectrum as a function of neutron multiplicity $F(m)$ equals the sum of events in the original spectrum $f(m)$ for multiplicities that are equal to or greater than $m$.

$$F(m) = f(m) + f(m+1) + f(m+2) + \ldots \qquad (1)$$

The advantage of handling integrated spectra is that there are no gaps in the data because $F(m) > 0$ for all multiplicities from 1 up to the highest recorded value. Additionally, statistical errors become less obtrusive as the number of events increases rapidly. The main disadvantage is that summing distorts the spectrum, smoothing out potential structures and shifting their peaks. For instance, if the original spectrum had a Gaussian peak on a flat background, the integrated spectrum would look like a smoothed step function reaching a maximum at about three sigma lower multiplicity. Nonetheless, spectrum integration (summing) remains a common and effective method for dealing with low statistics.

## 2.2 Smoothing

Instead of integrating spectra, one can use a smoothing procedure. Binning, which involves adding content from neighbouring channels, is the most common smoothing technique. This approach is effective when the number of channels surpasses the desired granularity. Another option is to average the contents of 3, 5, or 7 neighbouring channels. Unlike binning, this method does not reduce the spectra size; however, since the averaging window remains constant, the smoothing effect is strongest for small $m$-values (where it is unnecessary) and weakest for larger $m$-values.

Spallation, the dominant mechanism for muon-induced neutron generation, follows Poisson's statistics. Events with neutron multiplicity equal to $m$ are distributed within the $m \pm \sqrt{m}$ range. Therefore, the smoothing width should increase accordingly. This can be achieved, for example, by using Gaussian smoothing. In the resulting smoothed spectrum, events recorded with a multiplicity equal to $m$ are distributed among the neighbouring channels using a Gaussian distribution, with the $\sigma$ parameter (spread) increasing with the square root of multiplicity. Consequently, the number of events at each multiplicity will no longer be an integer, but the total number of events remains unchanged. Because the smoothing operation is independent (orthogonal) to the statistics, smoothing with $\sigma = \sqrt{m}$ produces a $\sigma = \sqrt{2}\sqrt{m}$ flattening of the spectrum. One may use $\sigma \leq \sqrt{m}$ to reduce the smoothing effect, provided it does not lead to artefacts such as multiple non-physical structures in the spectrum.

# 3 NMDS Setup and Measurements

NMDS stands for Neutron Multiplicity Detection System. The setup was constructed in the early 2000s at St. Petersburg's Khlopin Radium Institute by the team led by A.A. Rimsky-Korsakov and the University of Nevada at Las Vegas (UNLV) as part of the DOE Office of Nuclear Energy University Program. NMDS is schematically illustrated in Fig. 1. It comprises a 30 cm cube of lead (Pb) weighing 306 kg, surrounded by a 15 cm thick layer of high-density polyethylene (PE), which served as a moderator (Cestilene HD-500), with 60 embedded He-3 neutron counters operating in proportional mode and placed symmetrically around the target. Their positions were optimised to provide uniform detection efficiency for neutrons emanating from any point within the Pb cube. The active volume of each tube was a 28.5 cm long cylinder with a 1.55 cm diameter, filled with a 4 atm mixture of $^{3}$He (75%) and Ar (25%).

Each tube was fitted with a preamplifier and connected to a 1400 V high-voltage power supply. The computer-controlled high-voltage units were part of the data acquisition system that collected data to a standard PC. The digitiser had a time resolution of 1 μs and a 16-bit capacity. When a neutron was detected in any of the tubes, it was assigned to the time-bin #0, and the acquisition remained open for the registration of subsequent neutrons for the next 255 microseconds. After this period, the system was reset to prepare for the next event. The same neutron tube could register multiple events, provided they were spaced by 10 μs or more. Using the measured NMDS neutron lifetime of 65 μs, roughly 94% of neutrons from a single event were moderated to the thermal energy range (thermalised) within the 256 μs gate. Consequently, a 6% correction was applied during the analysis, reducing the efficiency for detecting multi-neutron events from 23.2±0.2%, as determined in experiments using a $^{252}$Cf source, to 22±2%. Measurements with the source and Monte Carlo (MC) simulations also confirmed that the detection efficiency is uniform throughout the Pb target volume. In some underground measurements, the system was enhanced with a 60 × 60 cm$^2$ shield of Geiger counters placed on the top of the setup, detecting the traversing charged particles.

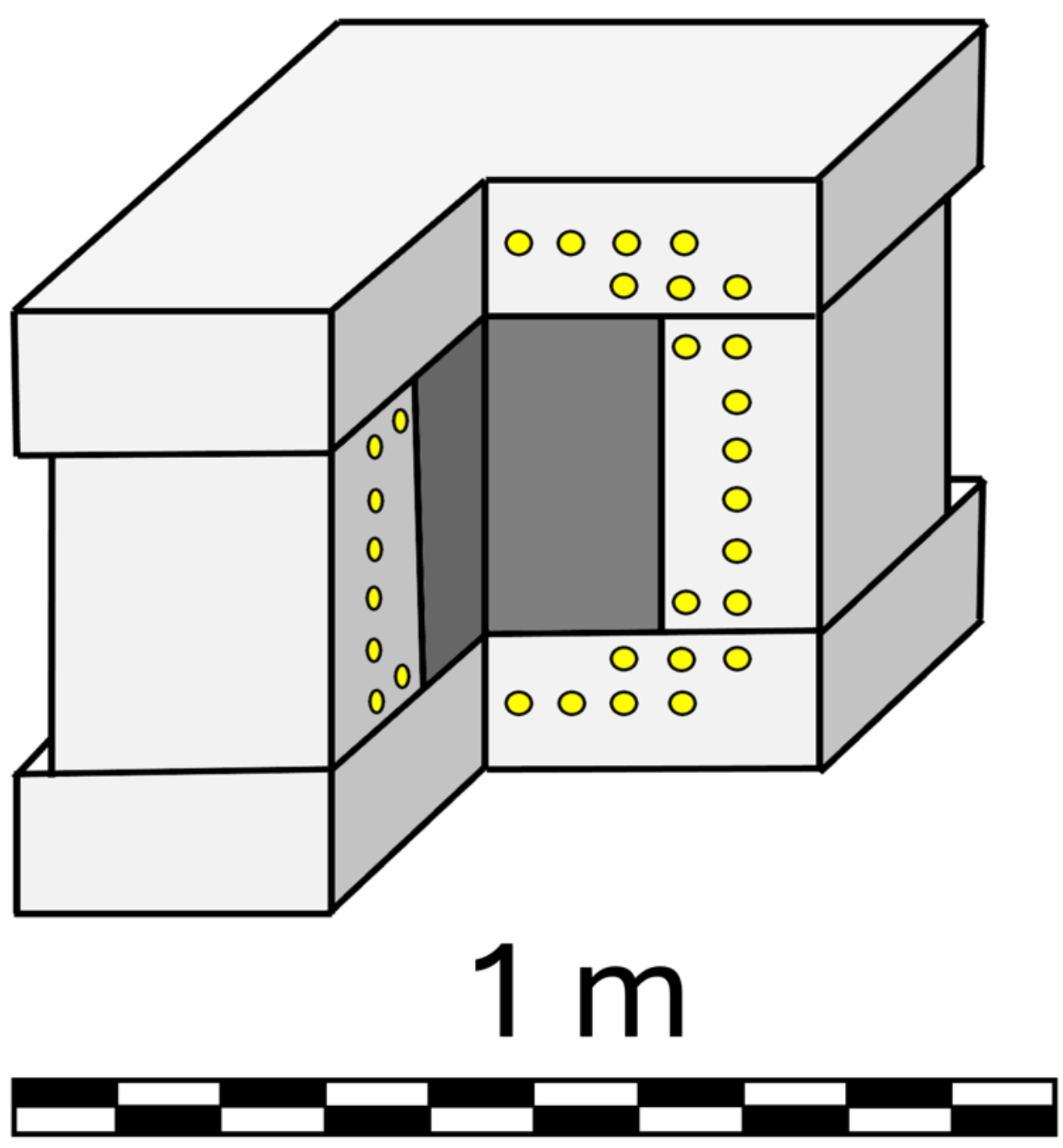

*Figure 1. Schematic representation of the NMDS setup: a 30 cm cube Pb target (dark grey) surrounded by a 15 cm polyethylene moderator (light grey) inlaid with 60 helium-3 neutron counters (yellow).*

Fig. 2 shows all integrated neutron multiplicity spectra measured using the NMDS setup. As the legend indicates, the data are derived from six observations at three different depths. At 583 m.w.e., we also measured neutrons in coincidence and in anticoincidence (vetoed) with a muon detector placed directly on top of NMDS. The vertical scale (neutron events per ton per month) has been chosen to account for variations in data-acquisition time. Throughout this paper, ton refers to a metric ton (1000 kg) and not a short ton (2000 lb).

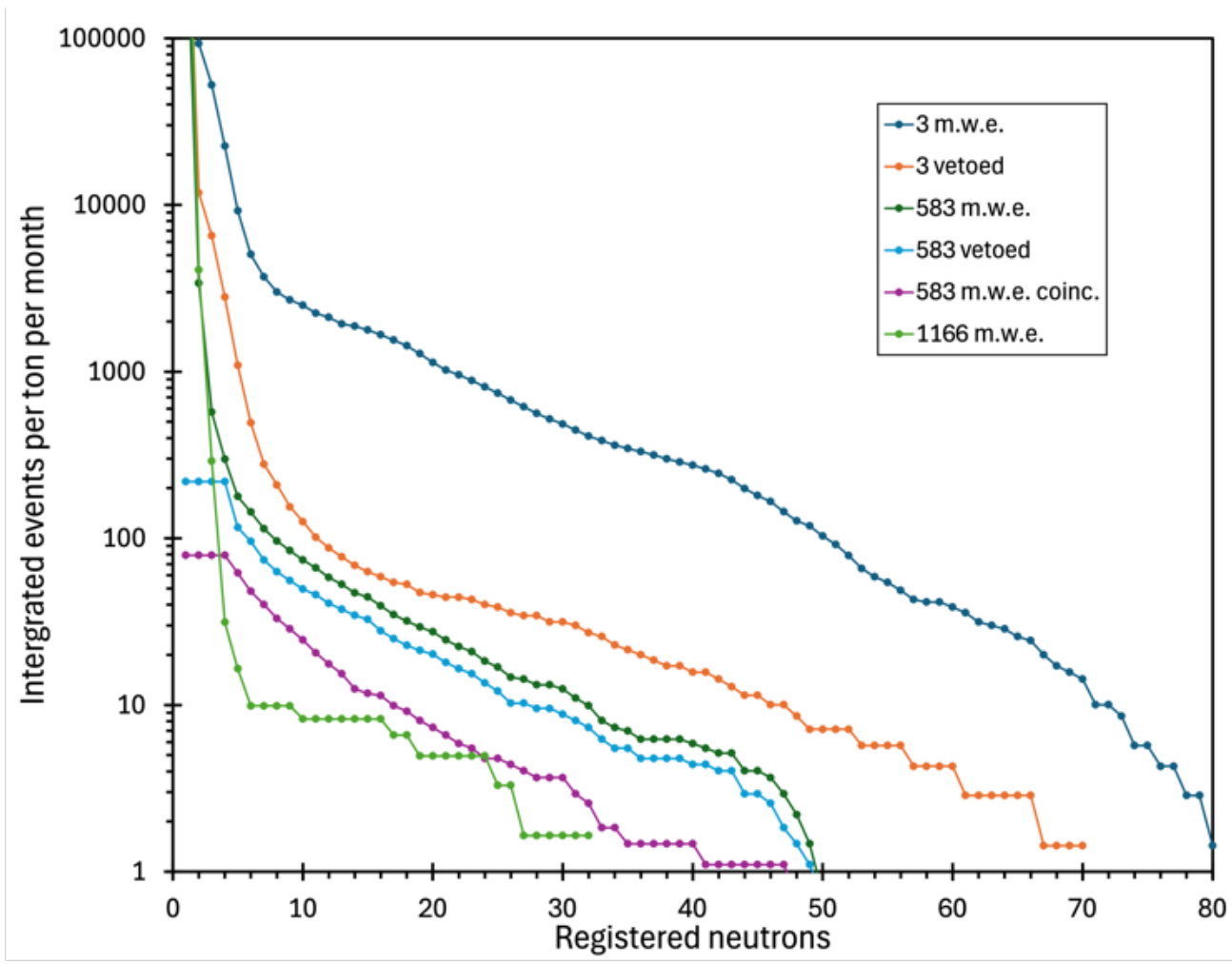

*Figure 2. Integrated neutron multiplicity spectra measured using the NMDS setup. The Y-axis displays events per metric ton per month to account for varying data acquisition times.*

## 3.1 Measurements at 440 m (1166 m.w.e.)

Measurements at the 440 m level in the Pyhäsalmi mine, corresponding to an overburden of 1166 m.w.e., commenced in September 2001. At that time, it was the deepest site accessible to our team. The measurements continued until January 2002, during which 1440 hours of data were acquired (0.612 ton-month). The results are shown in Fig. 3. Subsequent measurements [17] determined that the muon flux at that depth was 0.014 muons per square metre per second, that is, 1210 muons per square metre per day.

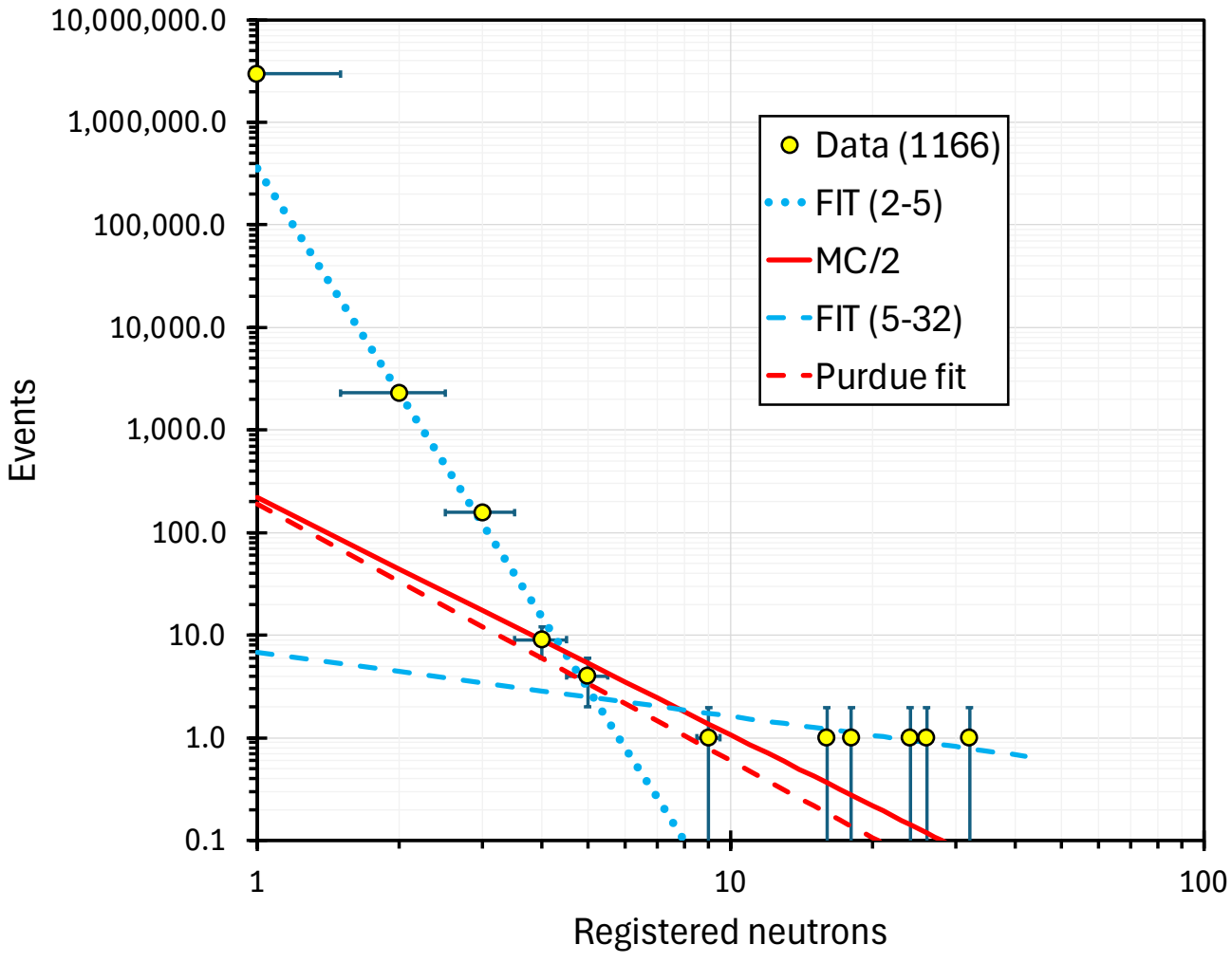

*Figure 3. Neutron multiplicity measured at 1166 m.w.e. (points with error bars). The blue dotted line is a power-law fit to the data with multiplicity $2 \leq m \leq 5$, and the blue dashed line, $m \geq 5$. The red line is the MC simulation adjusted by a factor of two, as explained in section 4.3. The red dashed line is the Purdue fit (Section 4.3).*

The two blue fits in Fig. 3 are intended to guide the eye and highlight that the two groups of events ($2 \leq m \leq 5$) and ($m \geq 5$) do not follow the MC trend (red line).

## 3.2 Measurements at 220 m (583 m.w.e.)

In February 2002, NMDS was relocated to a shallower site, offering better infrastructure, accessibility, and a higher muon count rate. At the same time, a Geiger counter array was added to the setup. It consisted of two layers of densely packed Geiger tubes, with an active volume of 60 × 60 × 10 cm$^3$, placed directly on the top wall. Although the coverage of this veto detector was only 14% of $4\pi$ (1.8 steradians), it was adequate to provide a sample of coincidence and anti-coincidence (vetoed) neutron multiplicity spectra.

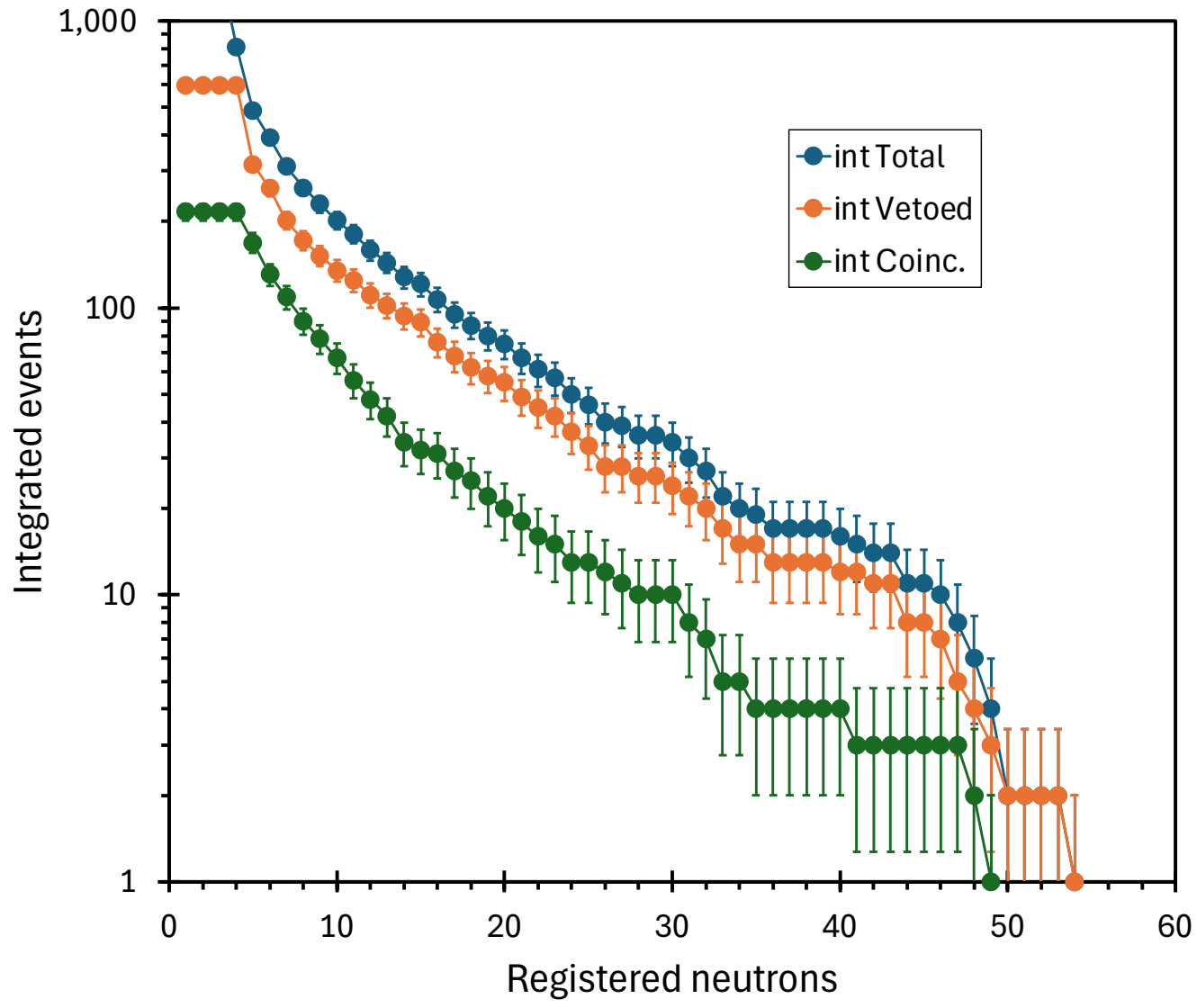

*Figure 4. Integrated total (blue points), vetoed (orange points), and coincidence (green points) neutron multiplicity measured at 583 m.w.e.*

Fig.4 shows the integrated muon multiplicity spectra taken at 583 m.w.e. The top series (blue) represents all the data, the middle (orange points) was taken in anti-coincidence (vetoed) with the Geiger array, and the bottom (green points) was in coincidence with the Geiger array. The spectra exhibit similar shapes. The average ratio between the total and vetoed points is 1.38±0.17 and remains nearly constant at all neutron multiplicities (Fig.5). It shows that the Geiger array introduces no significant multiplicity bias.

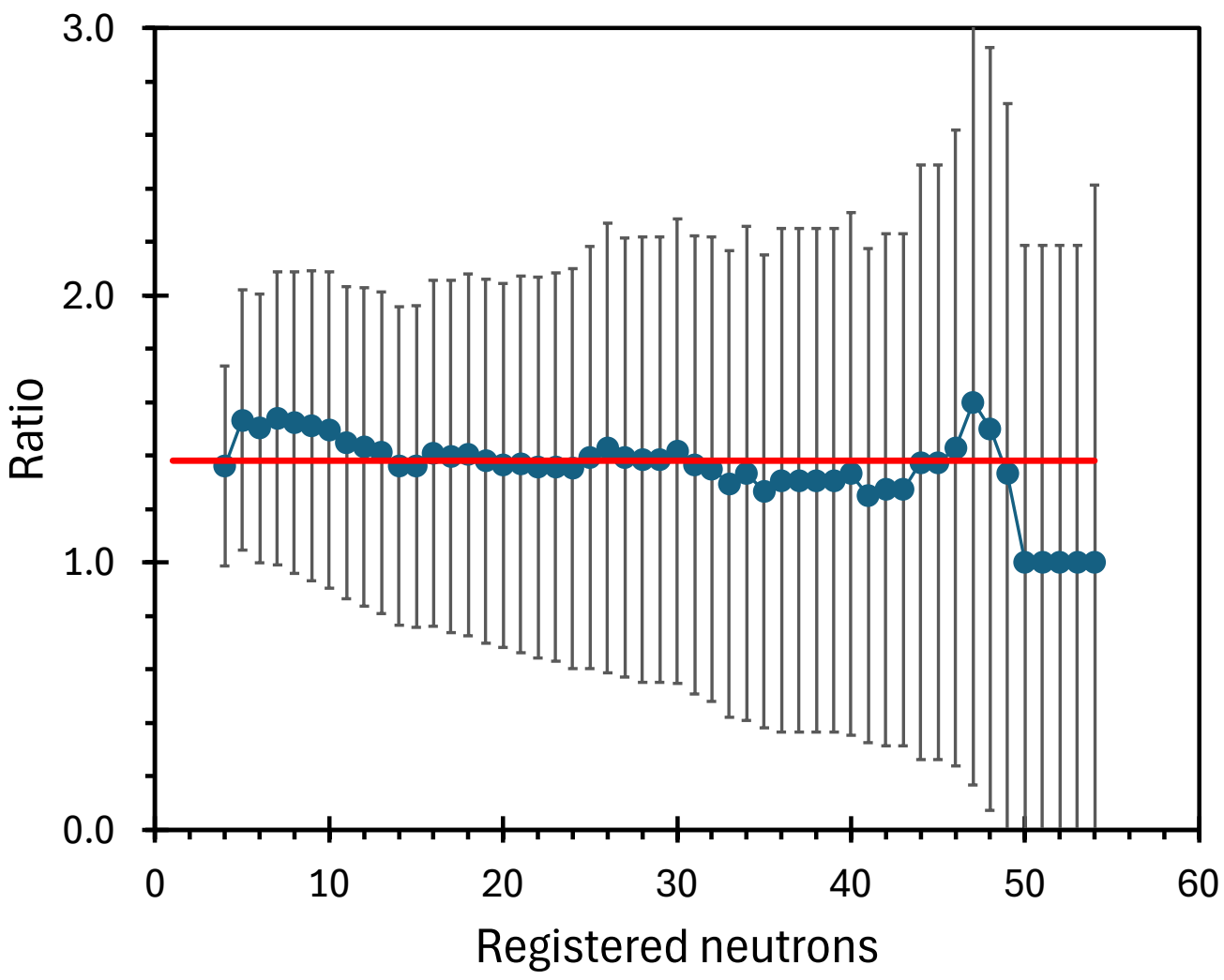

*Figure 5. The ratio of integrated total to vetoed neutron multiplicity spectra at 583 m.w.e. The solid red line indicates an average ratio of 1.38 ± 0.17.*

Spectra collected at 583 m.w.e. represent the best data in this compilation of six measurements conducted over the past two decades, utilising three experimental setups at different sites.

## 3.3 Measurements in the lab (3 m.w.e.)

The measurements were conducted in room #234 (one floor above the ground level) of the Khlopin Radium Institute building in St. Petersburg. Three metres of water equivalent correspond to approximately one metre of concrete, accounting for the roofs and ceilings of the floors above. Data were collected for 1,668 hours from May to July 2008 using a modified NMDS setup shown in the right panel of Fig. 6.

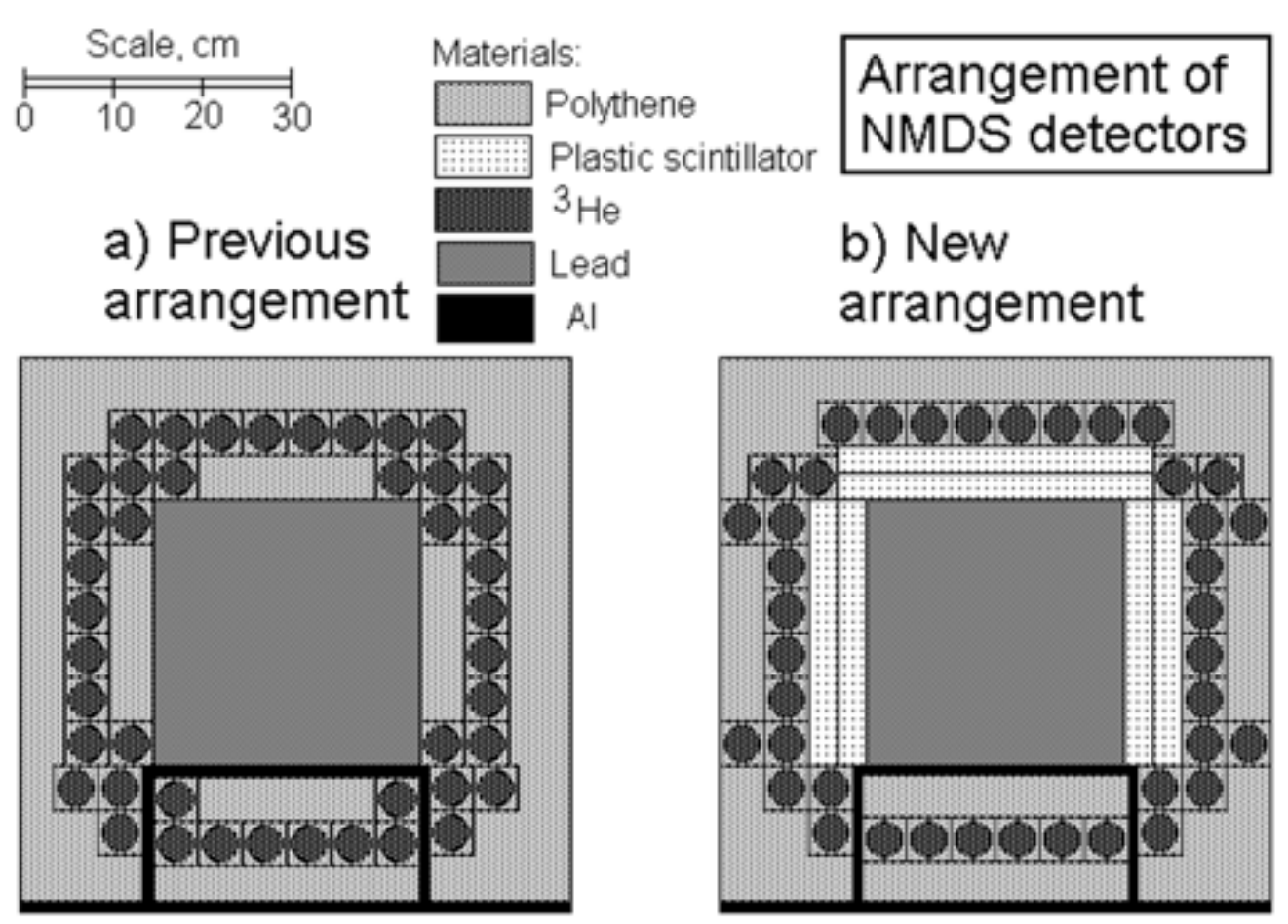

*Figure 6. Left: NMDS setup as used in the mine. Right: Modified NMDS setup.*

The modification involved 18 kg of a plastic scintillator, arranged as five detectors on the four sides and the top of the Pb cube (Fig. 6), providing ~80% of $4\pi$ coverage. Each detector comprised a sandwich of two scintillator plates measuring 30 × 30 × 2 cm³. Each plate was coupled with a Hamamatsu R6231-01 PMT. Adding five detectors in direct contact with the target necessitated a slight rearrangement of some of the 60 He-3 neutron counters and the removal of parts of the PE moderator. However, as the scintillator's moderating properties closely resemble PE's, the overall change in neutron detection efficiency was minimal. As checked with a $^{252}$Cf source, the modified NMDS retained 90% of the original setup's neutron detection efficiency.

A programmable logic controller processed the amplified and shaped signals from ten PMTs. Scintillator detectors are also sensitive to gamma rays. Their spectrum has a long exponential tail extending to high energies. To minimise gamma-induced signals, the acceptance threshold was set at 3 MeV, just below the signal's amplitude associated with charged particles (muons). To ensure compatibility with the old data format, the coincidence signal from the scintillator shield was added as input #0 among the 60 inputs designated for the neutron detectors. Naturally, some high-energy gamma rays could also generate a trigger. Since neutron detection did not accompany 92.4% of the scintillator triggers, data were recorded only if at least one neutron was present. The collected neutron multiplicity spectra are shown in Fig. 7.

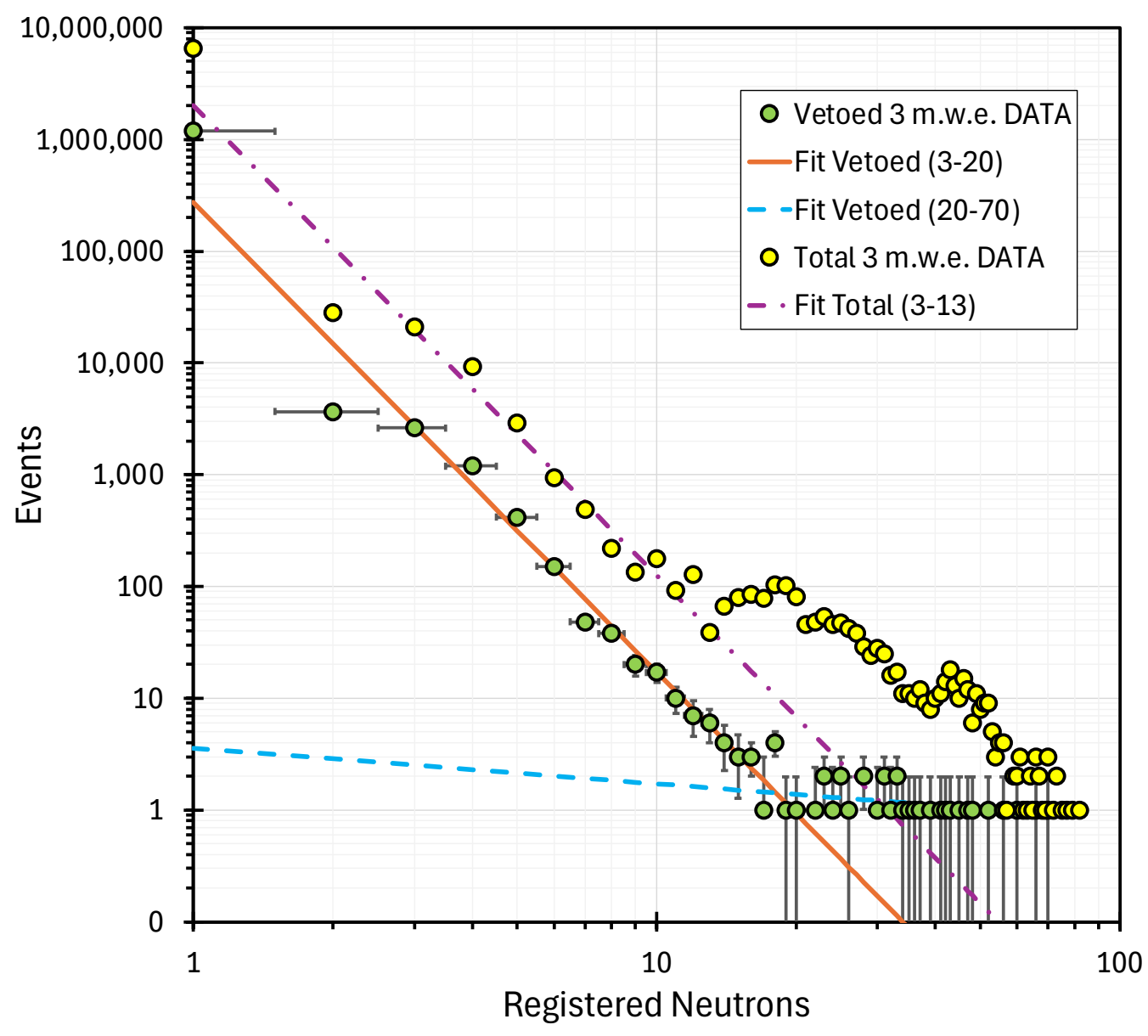

*Figure 7. NMDS data at 3 m.w.e. The yellow points are the Total spectrum. The green points with error bars are muon-suppressed (Vetoed) data. The lines are power-law fits to the data. The fitting ranges are indicated in the legend.*

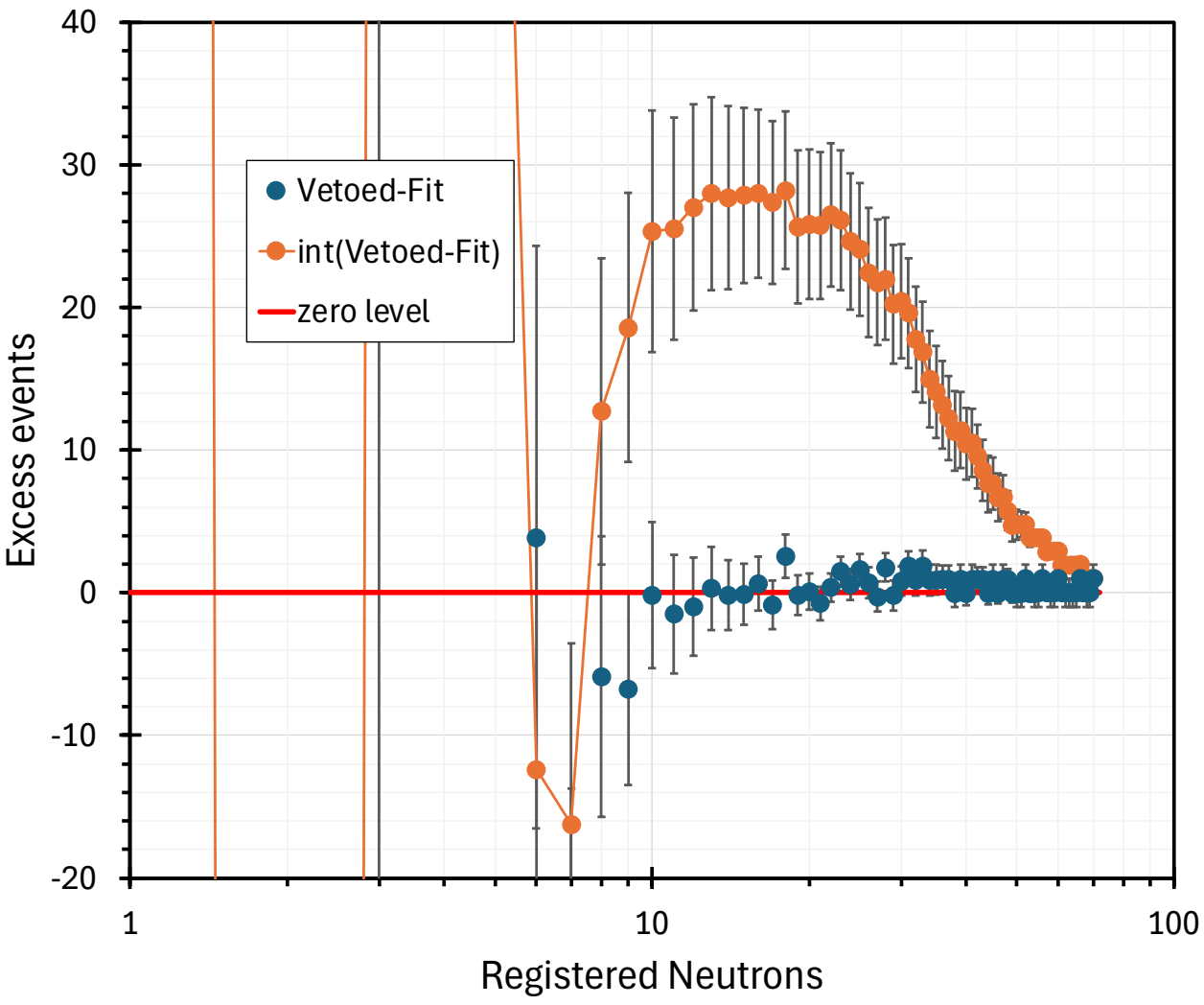

*Figure 8. Difference (blue points) and integrated difference (orange points) between the muon-suppressed (Vetoed) neutron events collected at 3 m.w.e. and the power-law fit (3-20), shown in Fig. 7.*

As explained in the Introduction, the instrumentation on the Earth's surface and at shallow locations is subject not only to muons but also to heavier charged particles originating from CR-induced showers. Therefore, we have searched for anomalies at 3 m.w.e in the vetoed neutron multiplicity spectrum rather than the total spectrum. Fig. 8 shows the excess events (blue points) remaining after subtracting the power-law function fitted to the data with multiplicity $3 \leq m \leq 20$, indicated in Fig. 7 as a solid red line. The integrated difference is the orange points in Fig. 8. It reaches a maximum at $m$=13 with 28$\pm$7 integrated excess events. Since the exposure was 0.7089 ton-month, the detected excess corresponds to 40$\pm$10 events per ton per month.

# 4 Evaluation of NMDS Spectra

## 4.1 Normalised NMDS Spectra

To identify and emphasise emerging trends from the NMDS measurements, we have replotted Fig. 2 by normalising the spectra at low neutron multiplicity. Due to the limitations of how the vetoed spectrum at 583 m.w.e. was acquired (with no data for $m < 4$), the lowest available normalisation point was $m = 4$. However, normalising at a slightly different neutron multiplicity would not fundamentally alter the overall picture. The outcome is illustrated in Fig. 9. Normalisation appears to have reversed the trend visible in Fig. 2, where the neutron multiplicity spectra progressively weakened with increasing overburden and declining muon flux. Normalising the spectra at $m = 4$ (Fig. 9) shows an apparent rise in the relative intensity of the mid-multiplicity ($6 < m < 46$) with the depth. When the registered multiplicity ($m$) is converted to the actual multiplicity ($M$), the range becomes $27 < M < 209$. For the conversion, the registered multiplicity ($m$) was divided by the 22% detection efficiency measured with a $^{252}Cf$ source and corrected for the fixed 256-microsecond data acquisition time gate, as explained in Section 3.

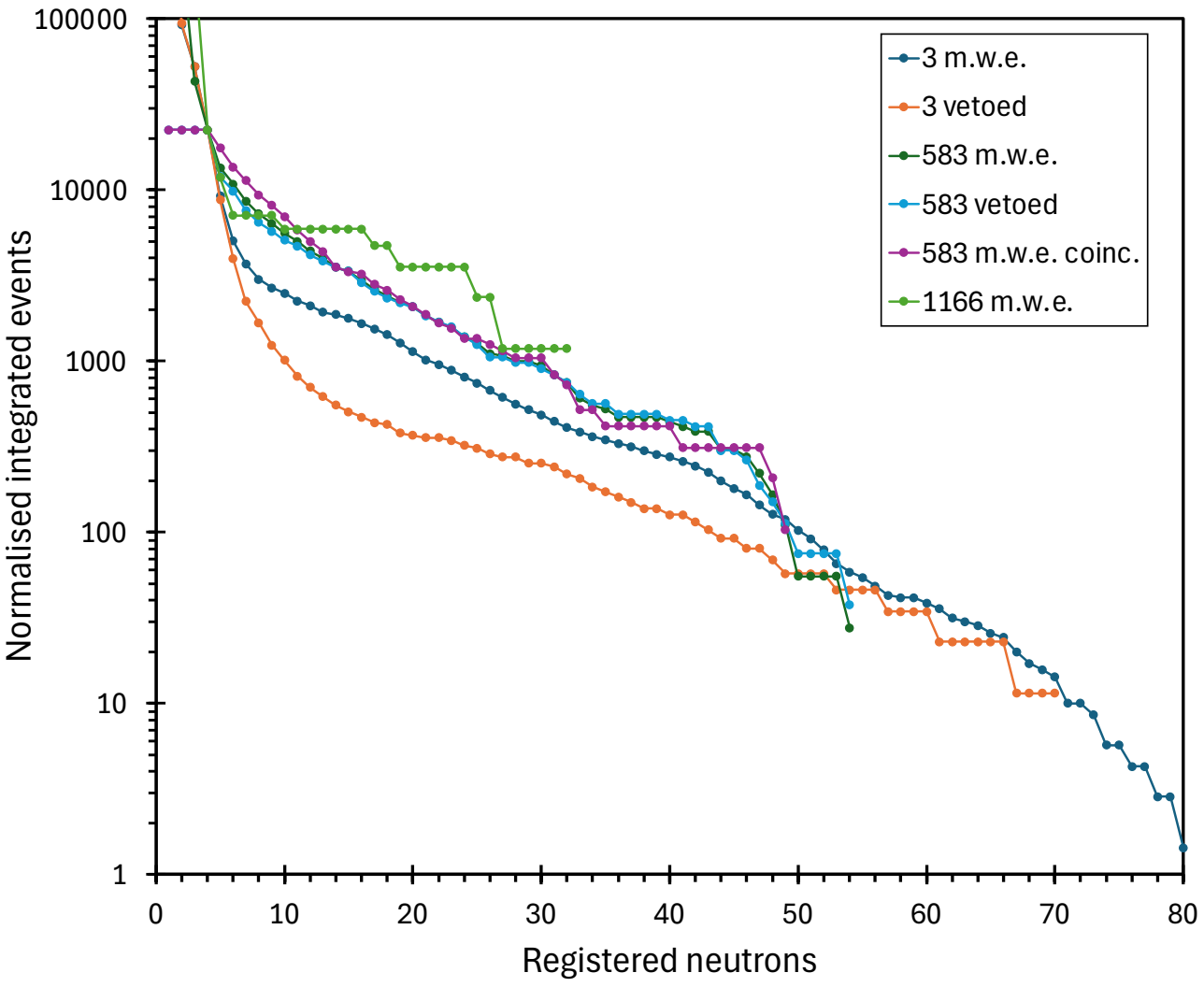

*Figure 9. NMDS data, normalised at multiplicity m=4.*

## 4.2 Normalised and linearised NMDS Spectra

As the number of registered events at any depth decreases by many orders of magnitude across low and high neutron multiplicities, linearised plots are a more convenient way to assess spectral shape changes. To do so, the data from Fig. 9 were multiplied by $m^{3.5}$ (neutron multiplicity to the power of 3.5) and renormalised in Fig. 10.

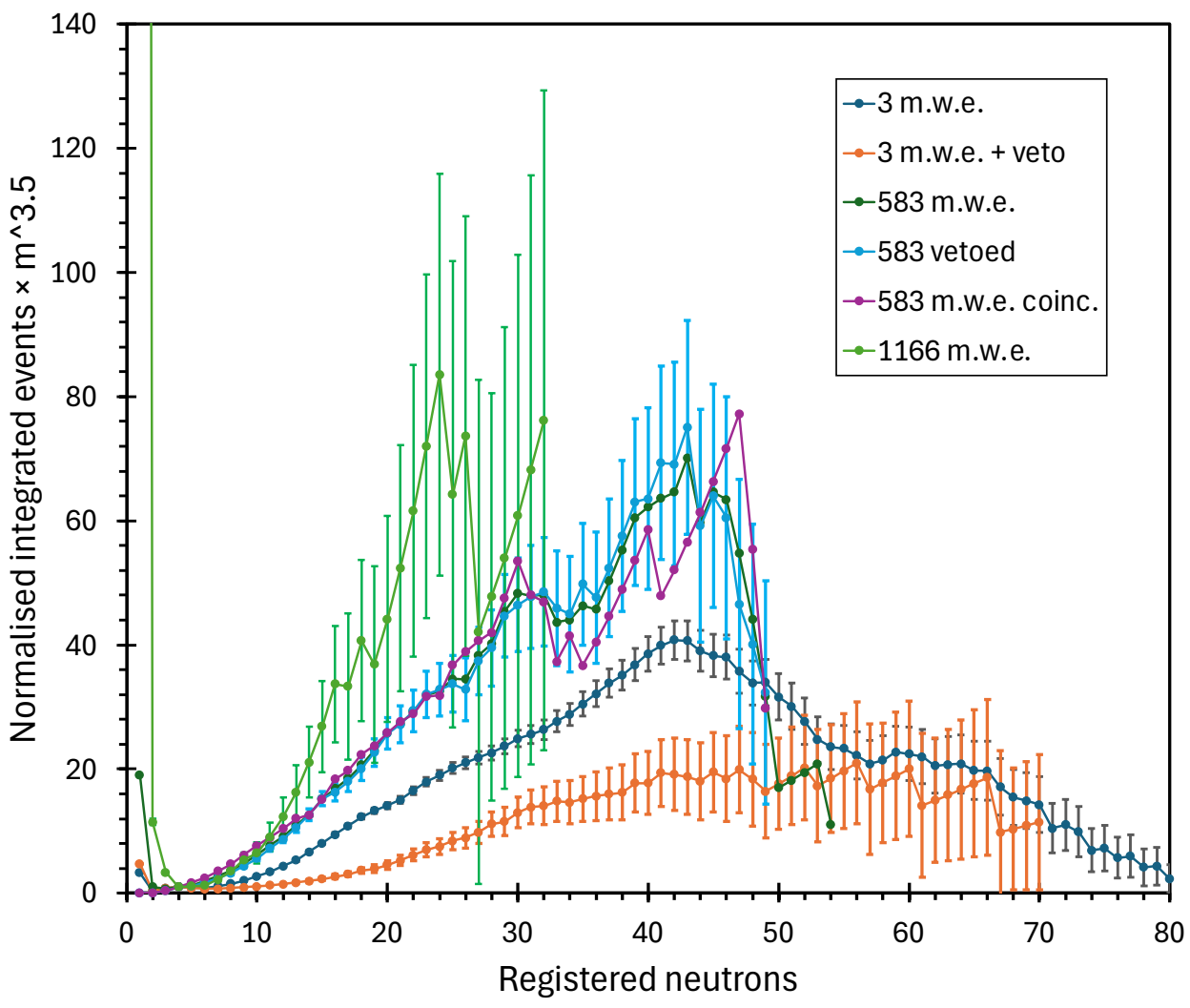

*Figure 10. Data from Fig. 9 multiplied by $m^{3.5}$ and normalised to unity at m=4. Such a transformation allows spectral shape comparison on a linear scale.*

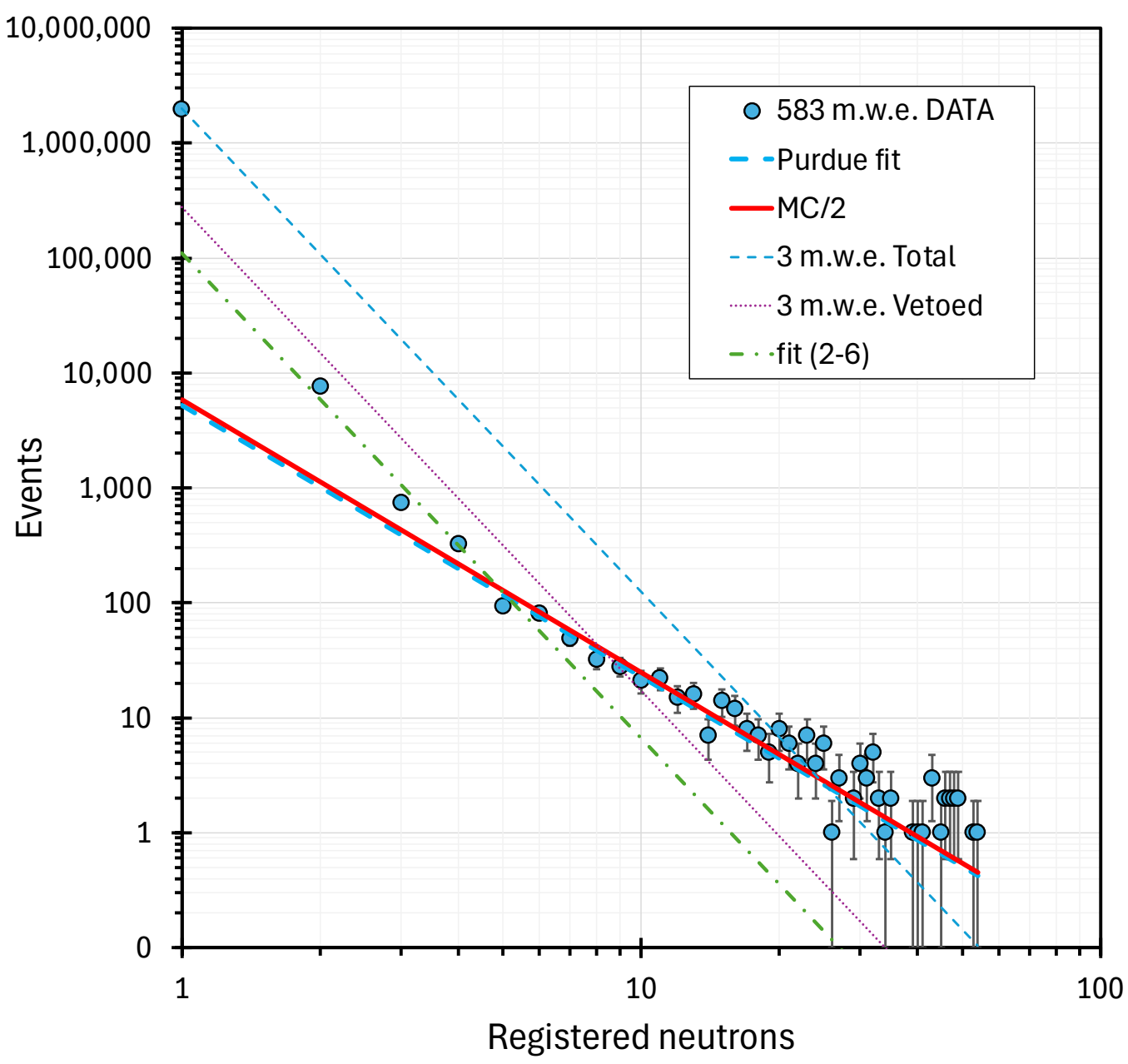

*Figure 11. Total 583 m.w.e. spectrum (points), the Purdue fit to the data (blue dashed line), and the MC simulation reduced by 50% (red line), as explained in section 4.3. The fit and the normalised MC are practically identical. The thin blue dashed and purple dotted lines are the trends for 3 m.w.e. data from Fig. 7. The green dash-dotted line is the outcome of a power-law fit to 583 m.w.e. data points $2 \leq m \leq 6$. The resulting slope matches the 3 m.w.e. trends but distinctly differs from MC and the Purdue fit.*

## 4.3 NMDS Monte Carlo Simulations

Significant effort and resources were devoted to simulating muon-induced neutron generation in and around the NMDS setup at 583 and 1166 m.w.e. and their subsequent detection in the 60 He-3 counters. These simulations were the focus of a PhD Dissertation defended by Haichuan Cao in May 2023 at the Department of Physics and Astronomy at Purdue University, Indiana, USA [18] and summarised in an arXiv paper [7].

The primary outcome of this dissertation is that Geant4 simulations of the registered neutron multiplicity spectra are well-approximated by a two-parameter power-law function $k \times m^{-p}$, where $k$ and $p$ are the parameters and $m$ is the registered neutron multiplicity. While the simulation accurately determines the slope parameter $p$, the Geant4-predicted amplitude parameter $k$ is twice the value fitted to the experimental data. Therefore, to achieve good agreement, the amplitude must be extracted from the fit or reduced by 50%. It was also concluded that the slope parameters $p$, measured and simulated, at 583 m.w.e. and 1166 m.w.e. are practically the same. The results for 583 m.w.e. are: $p_{MC}$ = 2.37±0.01 and $p_{fit}$ = 2.36±0.10. For 1166 m.w.e., the corresponding values are: $p_{MC}$ = 2.3±0.01 and $p_{fit}$ = 2.50±0.35. The amplitude parameters (in thousands) are $k_{MC}$ = 11.6±0.28 and $k_{fit}$ = 5.2±1.2 for 583 m.w.e., and $k_{MC}$ = 0.44±0.11 and $k_{fit}$ = 0.19±0.17 for 1166 m.w.e. The quoted errors are statistical only. Systematic errors are not quantified, but neither this nor any other MC simulation predicts two-sloped spectra like the one shown in Fig. 3. Still, at this point, it is unclear whether we are dealing with an omission of some interactions in the simulations, an experimental artefact, or a genuinely new phenomenon.

Fig. 11 shows the NMDS total spectrum measured at 583 m.w.e. (points with error bars). The thick blue dashed line is the Purdue power-law fit ($p$=2.36, $k$=5200). It practically overlaps with the fit to the MC outcome (solid red line). We treat the fitted line as a reliable parametrisation of the muon-induced neutron contribution and subtract it from the measured spectrum. If anomalies are present, they would manifest themselves as a statistically significant excess in the subtracted spectrum. Otherwise, the resulting points would be evenly distributed on both sides of the zero line. Integrating the background-subtracted spectra may further enhance the visibility and strengthen the evidence for anomalies. In Fig. 12, the fitted function ($p$=2.36, $k$=5200) was subtracted from the data (yellow points). The blue points represent the integrated difference between the data and the fit. Although the yellow points seem to oscillate around zero, as they should, the integrated values (blue points) cumulate and reveal an excess of events for multiplicities $5 < m < 50$, peaking at $m \sim 11$ with 42±14 events, i.e. 15±5 events per (metric) ton-month; an apparent three-sigma effect.

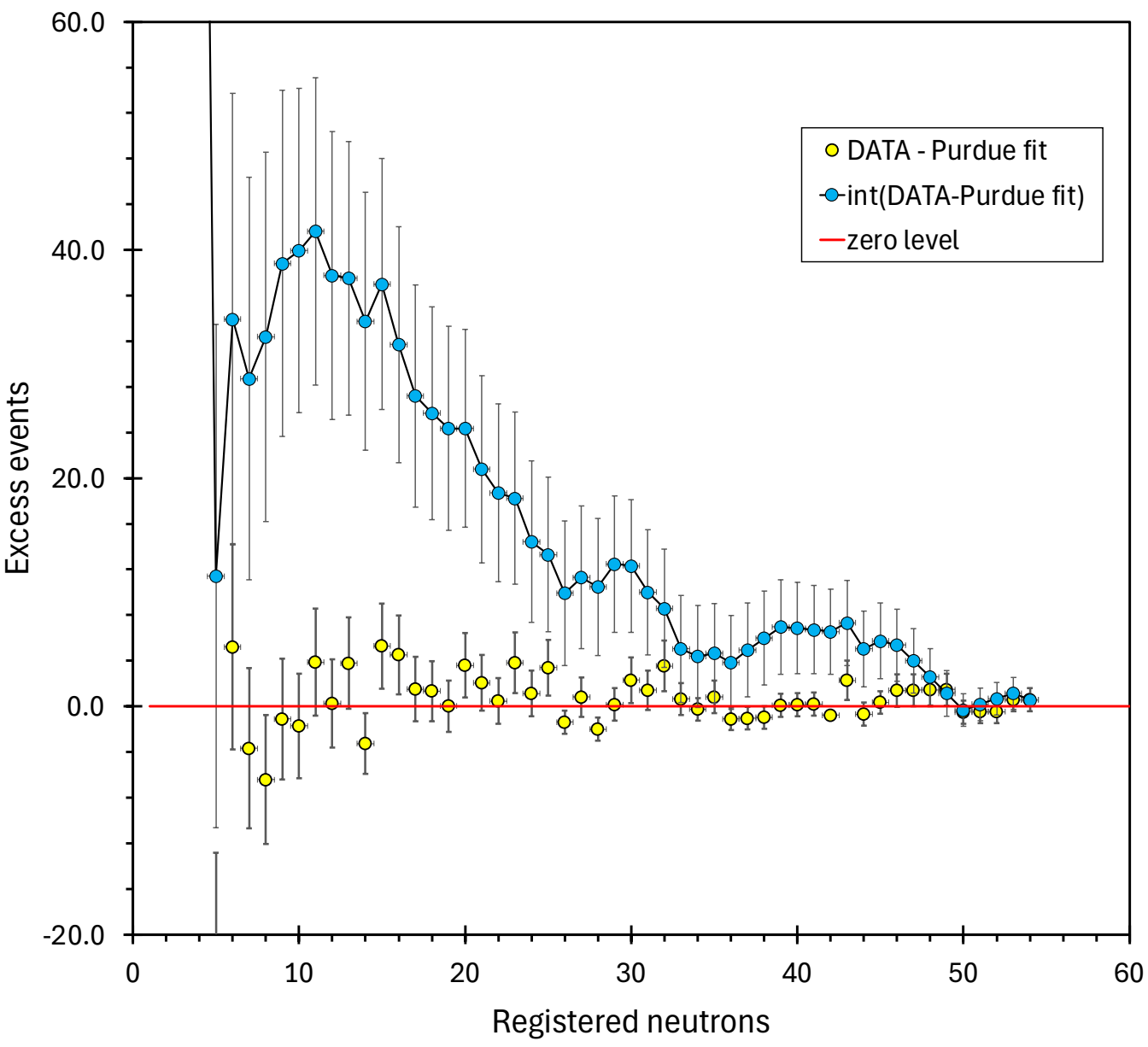

*Figure 12. Yellow points: NMDS 583 m.w.e. data after Purdue power-law fit subtraction. Blue points: the same data after integration.*

To incorporate the fitted function and the MC-simulated trends into Fig. 10, integrated and normalised values of these power-law dependencies are needed. Integrating data is noncontroversial. Integration ends with the last channel containing data. However, it is not clear how to handle power-law functions. The problem is illustrated in Fig. 13. Integrated experimental data points are shown as dots with error bars. The top line shows the outcome of the integration of the Purdue fit up to multiplicity 200; the middle, up to $m$ = 64 (ten channels above the highest recorded event), and the bottom, up to $m$ = 54 (the highest recorded event). None of the trends fully represents the measured data. Still, the middle red curve provides the best match, and it is used (for visualisation only) in Fig. 14. Otherwise, the integrated power-law functions are not needed for the analysis. The quantitative values are extracted by integrating the difference between the measured data and the power-law function; therefore, the integration range ambiguity is resolved.

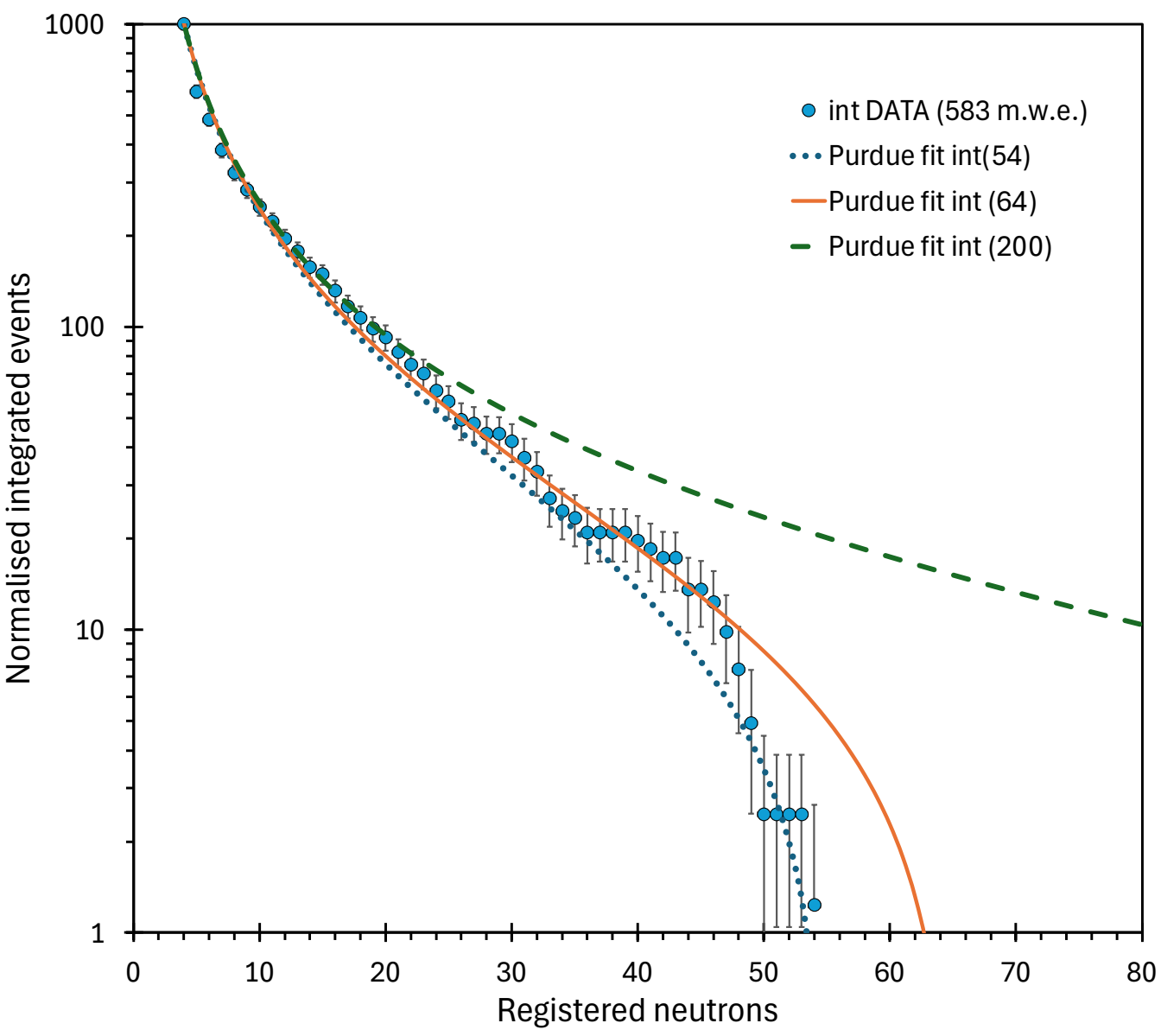

*Figure 13. The top dashed line is the outcome of integrating the Purdue power-law fit up to multiplicity 200; the middle (red solid line), up to m = 64, and the bottom (blue dotted line), up to m = 54. The 583 m.w.e. data points are also shown.*

Fig. 14 compares the measured values at 3 m.w.e (red points), 583 m.w.e. (orange points) and at 1166 m.w.e (green points) with the Purdue fit to 583 m.w.e. data (green solid line) and MC predictions for 1166 m.w.e. depth (black solid line) and 583 m.w.e. (black dashed line). The red dotted lines in Fig. 14 represent the shapes of the 3 m.w.e. data (read points) multiplied by 1.7 and 3.0, to match the measured spectra at 583 and 1166 m.w.e., respectively. As the slope parameters of the Purdue fit ($p$=2.36) and the MC prediction ($p$=2.37) are similar, and both were integrated up to m=64, the two lines are very close in Fig. 14 and follow the 583 m.w.e. data trend up to $m$=47. However, the measured 1166 m.w.e. values (green points) deviate noticeably from the MC-expected trend (black solid line) integrated up to $m$=74. The 1166 m.w.e. data ends abruptly at $m$=32 due to low statistics (Fig. 3). The acquisition time should have been significantly longer to cover higher multiplicities.

Another feature of the 1166 m.w.e. spectrum (Fig. 3) is the discrepancy between the Purdue fit, the MC, and the data. Two different power-law fits are needed to describe the measured values: for $2 \leq m \leq 5$ (blue dotted line) and $5 \leq m \leq 32$ (blue dashed line). The former is considerably steeper ($p$=7.26), and the latter is significantly flatter ($p$=0.623) than the MC prediction (p=2.31). The inadequacy of the MC prediction for 1166 m.w.e. is also visible in Fig. 14.

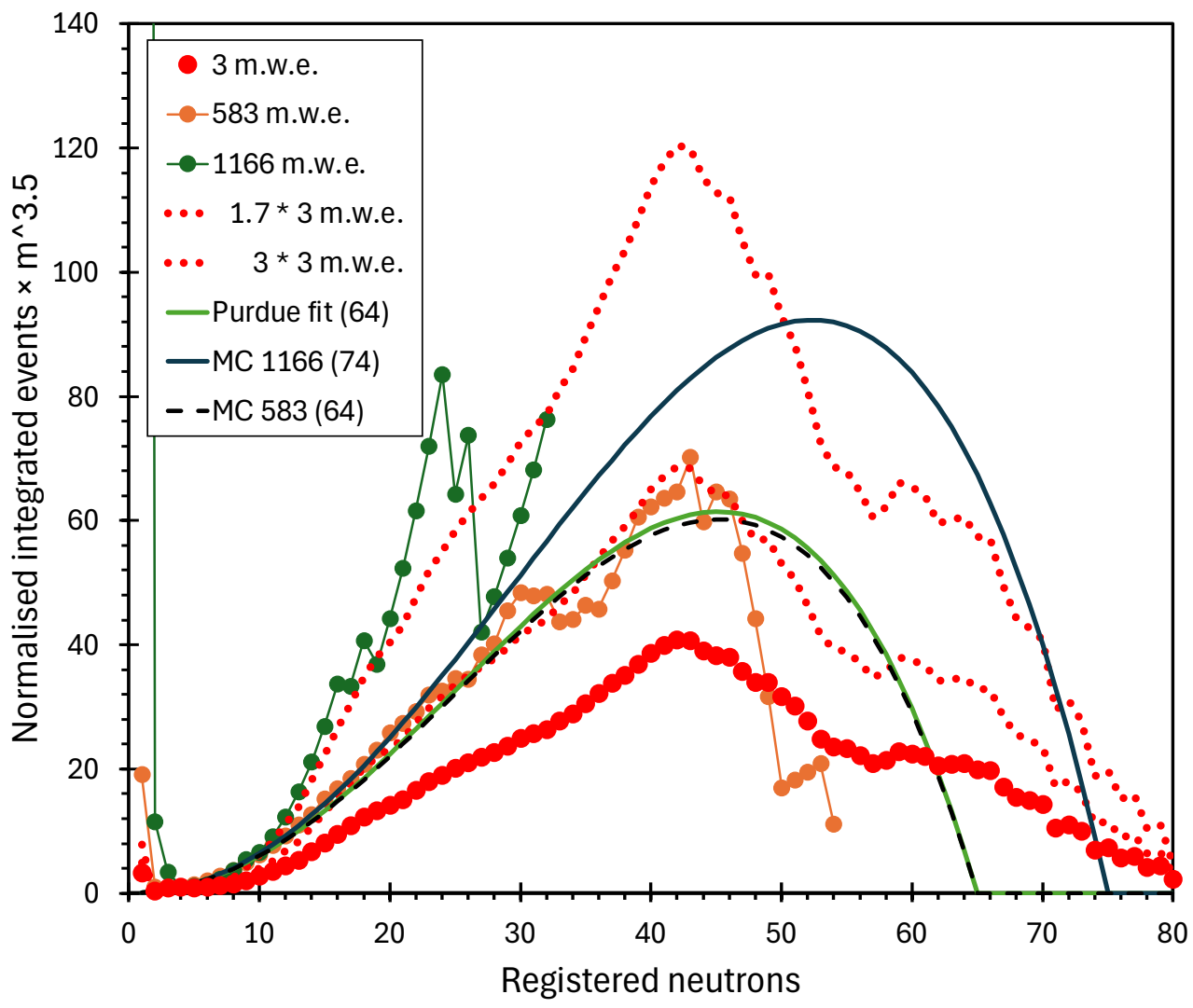

*Figure 14. The 3, 583, and 1166 m.w.e. data points are from Fig. 10. The thin red dotted lines show the 3 m.w.e. spectrum (red open points) multiplied by 1.7 and 3.0 to match the shapes of the normalised spectra at 583 and 1166 m.w.e., respectively. The solid green curve is the integrated Purdue fit. The black curves are the integrated MC predictions for 583 (dashed) and 1166 m.w.e. (solid).*

## 4.4 Coincidence and Anticoincidence Neutron Multiplicity Spectra

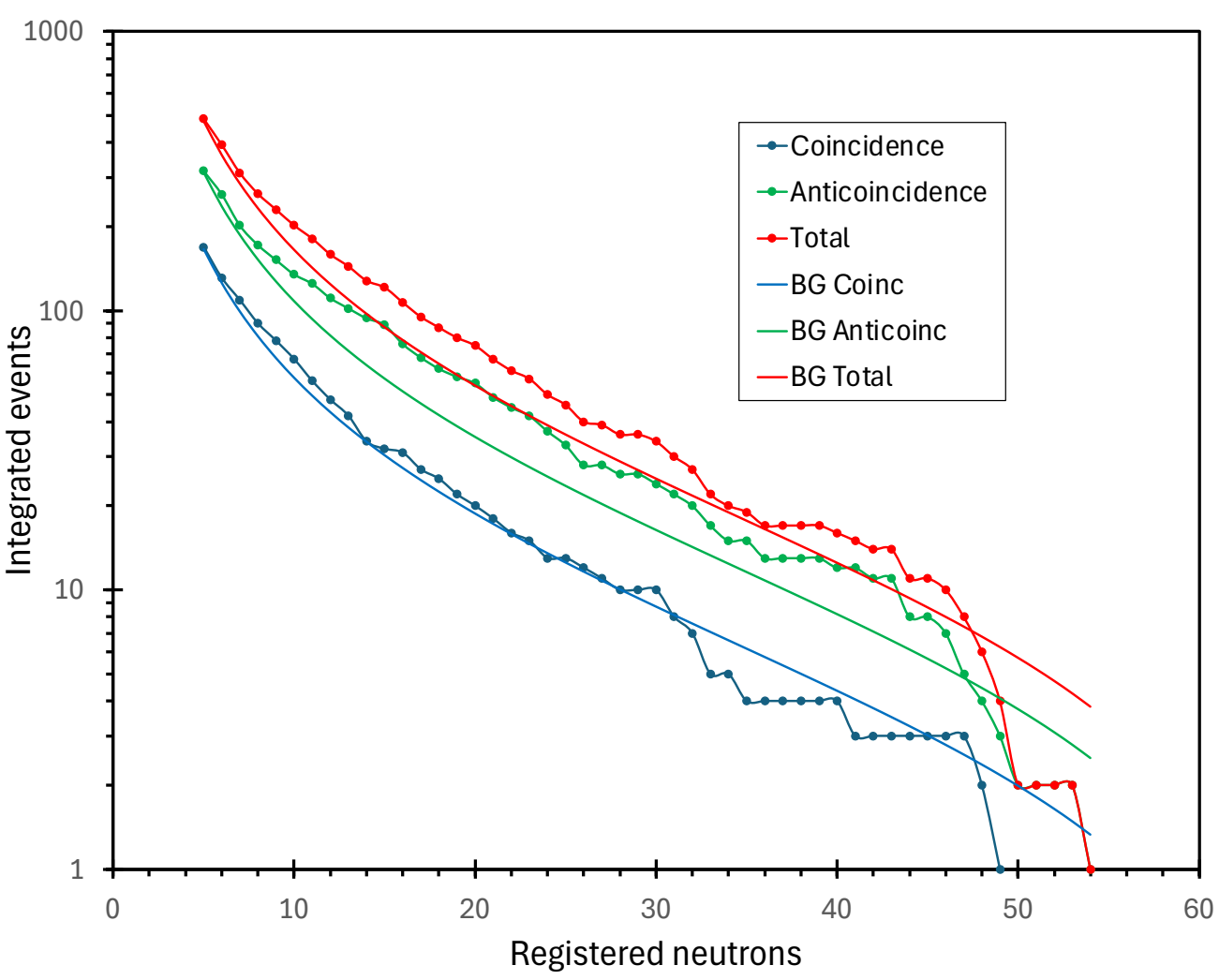

*Figure 15. Integrated total (red points), anticoincidence (green points), coincidence (blue points) data and power-law parametrised muon-induced neutron multiplicity background spectra.*

In addition to the total spectrum, coincidence and anticoincidence (vetoed) spectra were also recorded at 583 m.w.e. The Purdue MC study evaluated the power-law parameters only for the Total neutron multiplicity spectrum. While the shape parameter stays the same ($p$=2.36), determining the amplitude parameters for the coincidence and anticoincidence spectra was needed. Fig. 15 illustrates our $k$-finding procedure. As the integrated total spectrum and its power-law background cross at $m$=5, we adjusted the k parameters of the coincidence and anticoincidence muon-induced background spectra to match this property. The resulting parameters are $k$=1808 for the former and $k$=3392 for the latter.

If the anomalous excess in neutron multiplicity is genuine and unrelated to the muon flux, one would expect a diminished effect for the spectrum recorded in coincidence with the muon-triggered Geiger array. At the same time, the anti-coincidence spectrum would retain the excess and be less contaminated with muon-induced neutrons at the cost of somewhat lower statistics. This is indeed the case, as shown in Fig.16. While the excess in the total spectrum is 42±14 events and 34±11 events in the anticoincidence spectrum, there are at most 12±11 excess events in the coincidence spectrum. These values correspond to 15±5, 13±4, and 4.3±3.9 events per metric ton per month, respectively.

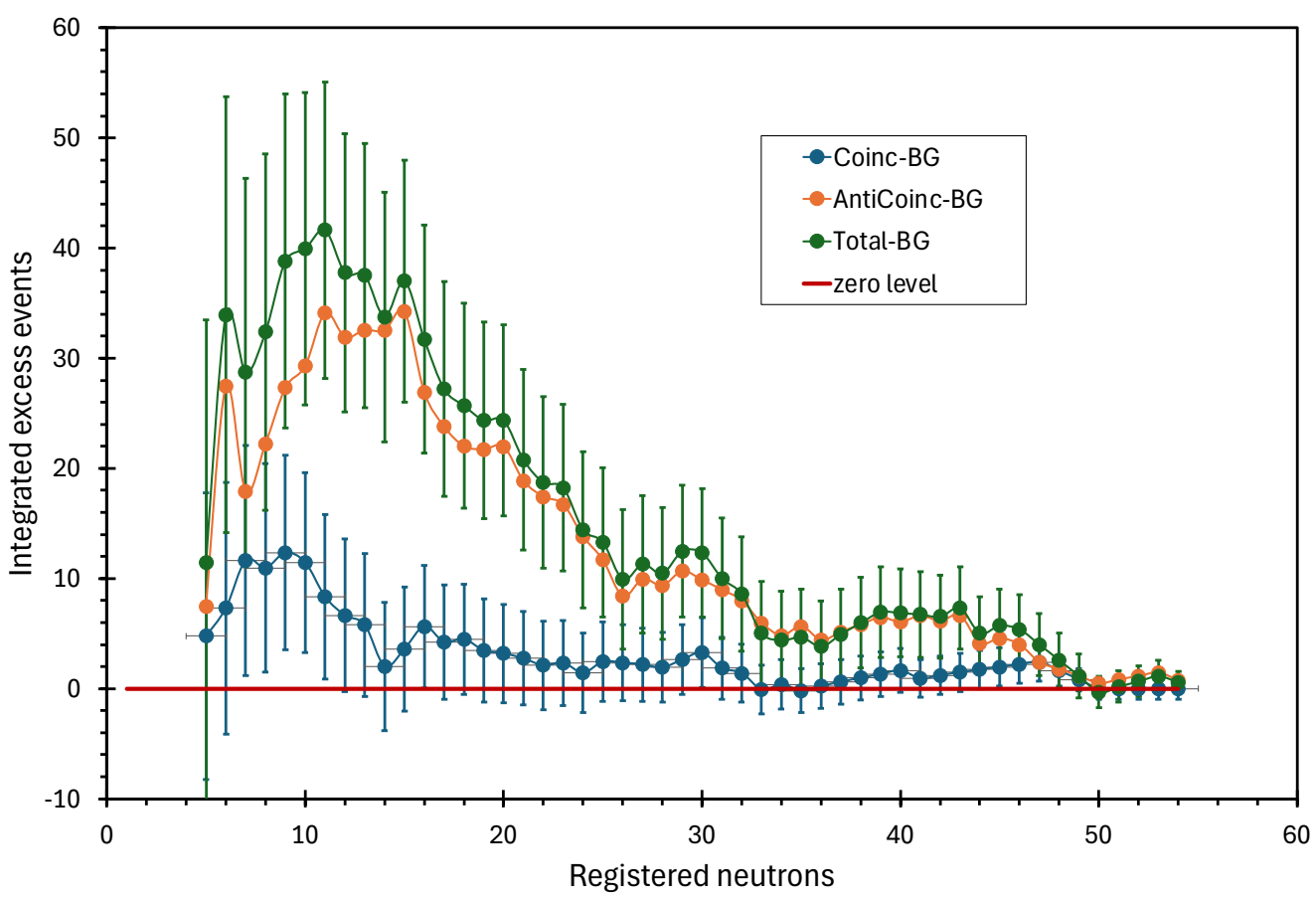

*Figure 16. Integrated excess events in the background-subtracted coincidence, anticoincidence, and total neutron multiplicity spectra. For clarity, statistical error bars for the anticoincidence points are omitted as they are practically the same as those for the total spectrum.*

## 4.5 Smoothed Neutron Multiplicity Spectra

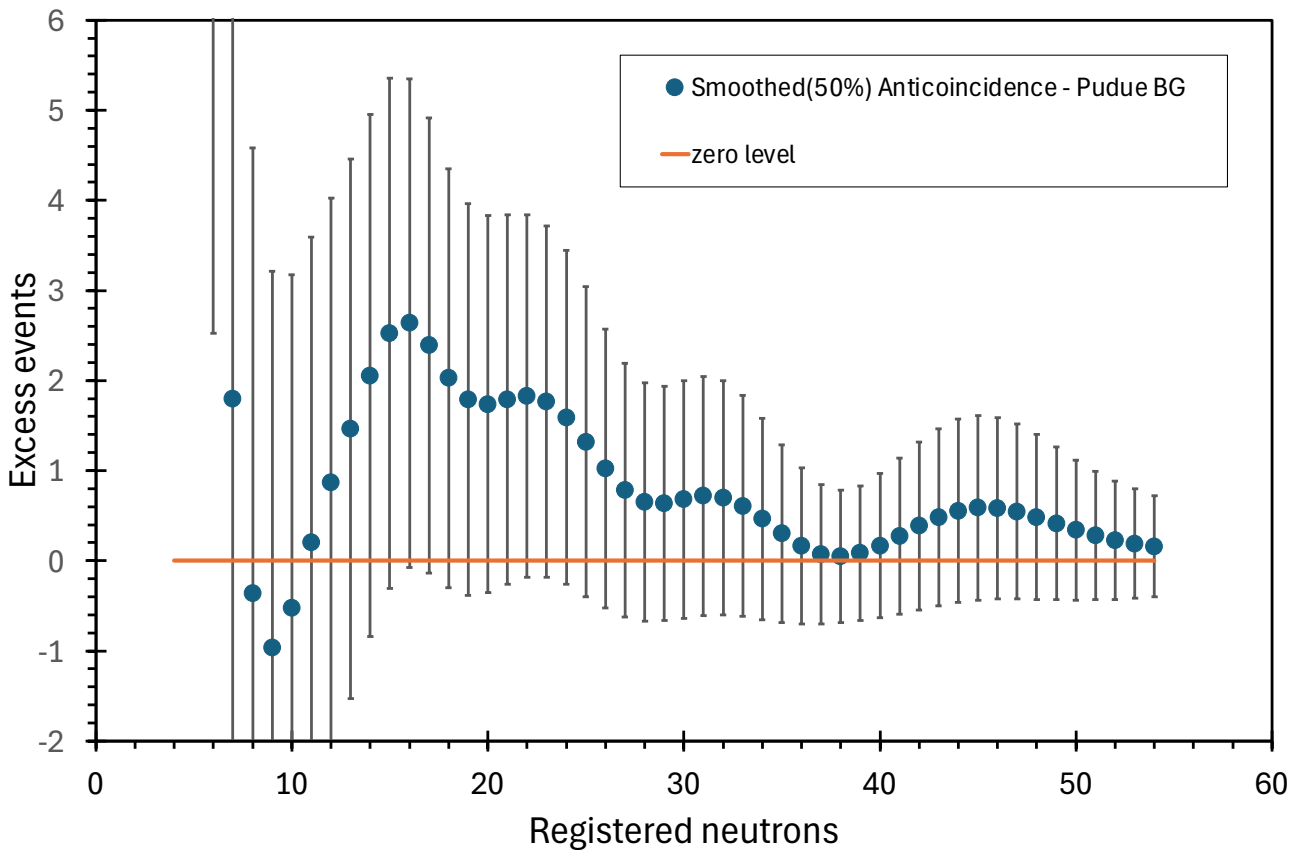

*Figure 17. Background-subtracted Anticoincidence neutron multiplicity spectrum smoothed with* $\sigma = 0.5\sqrt{m}$.

As explained in Section 2.2, smoothing is a useful tool for addressing low statistics. Fig. 17 displays the smoothed, non-integrated, background-subtracted neutron multiplicity spectrum obtained in anticoincidence with the Geiger array at 583 m.w.e. Regarding muon contamination, it is the cleanest high-statistics data sample we have. To minimise smearing of possible components of the multiplicity spectrum, the smoothing was performed with $\sigma = 0.5\sqrt{m}$ a width equal to half the statistical spread. The outcome has a structure resembling four Gaussian peaks. While the existence of the excess events in the range of $10 < m < 54$ is evident as all the points are above the zero level (red line in Fig. 17), the available statistics do not warrant a conclusive statement on the shape and structure of the excess events. Indeed, smoothing procedures tend to induce slight undulations. To check if this is the case, we have repeated smoothing with $\sigma$ = 40, 50, and 100% $\sqrt{m}$ and fitted Gaussian peaks to the resulting spectra. If the peaks were artefacts, their number, positions, and areas would significantly change or disappear with intensified smoothing. If not, they may be real. All four peaks survived the test. Their position remained within one multiplicity channel, and the peak areas stayed within 2% for peaks 0, 2, and 3 and 19% for peak 1. Fig. 18 shows the four peaks fitted to the spectra smoothed with $\sigma = 40\%\sqrt{m}$ (dotted lines), $\sigma = 50\%\sqrt{m}$ (dashed lines), and $\sigma = 100\%\sqrt{m}$ (solid lines). The black solid line is the $\sigma = 100\%\sqrt{m}$ smoothed background-subtracted spectrum, and the red points are the fitting residue. Given the minimal residue, the four-peak description matches the data. The parameters of the four peaks, obtained by fitting to the $\sigma = 100\%\sqrt{m}$ smoothed spectrum, are in Table 1. The peaks are also visible in the smoothed, background-subtracted Total 583 m.w.e. spectrum, which is nearly identical to that shown in Fig. 17.

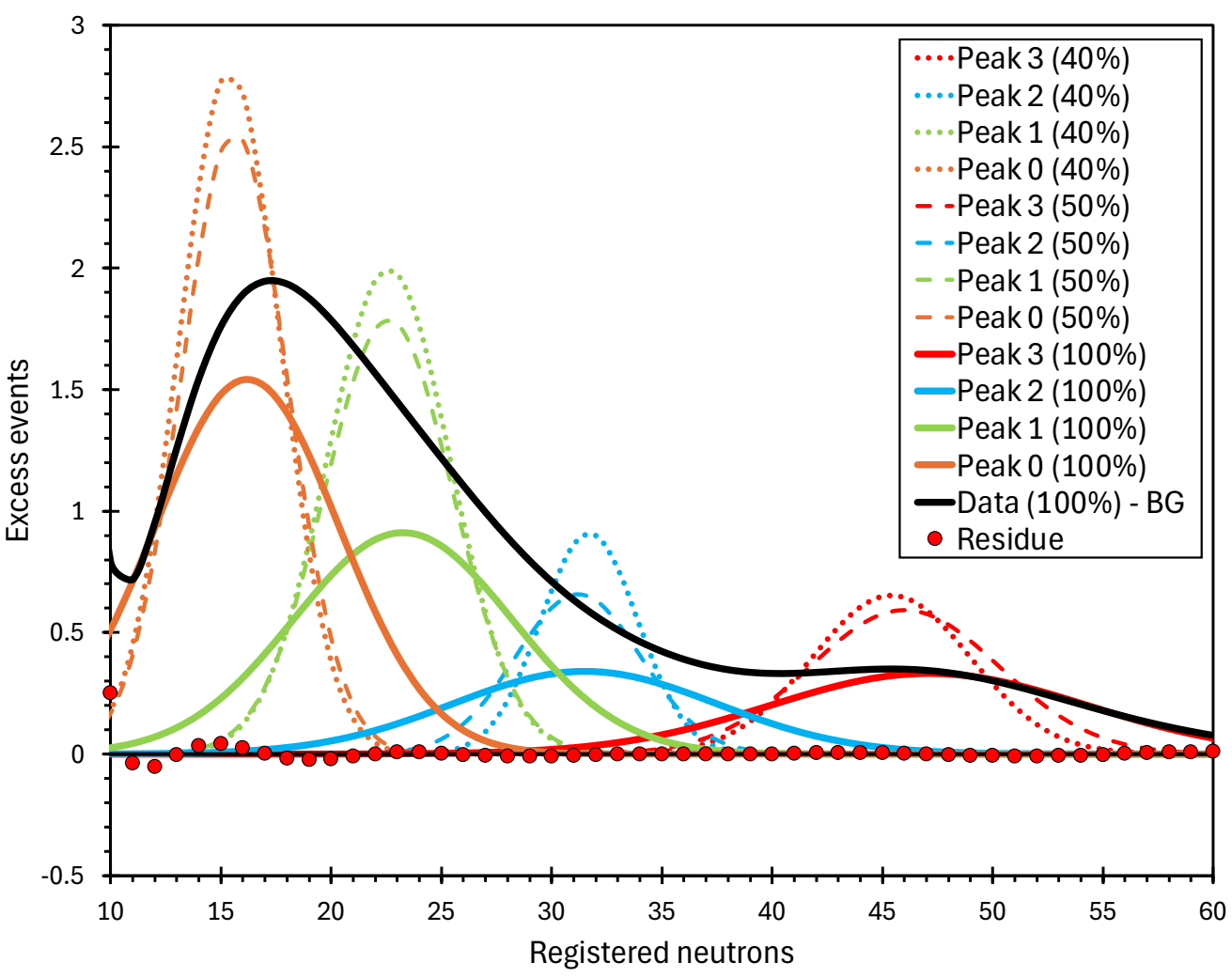

*Figure 18. The four peaks fitted to the spectra smoothed with $\sigma = 40\%\sqrt{m}$ (dotted lines), $\sigma = 50\%\sqrt{m}$ (dashed lines), and $\sigma = 100\%\sqrt{m}$ (solid lines). The black solid line is the $\sigma = 100\%\sqrt{m}$ smoothed background-subtracted spectrum, and the red points are the fitting residue.*

Since the four peaks retain their positions and areas under progressive smoothing, their width ($\sigma$) parameters positively correlate with peak positions and are greater or equal to the square root of their position, as expected from the statistics and our neutron detection efficiency, we treat the fitted peaks as a plausible hypothesis to be verified by future measurements. If the energy spectrum of the excess neutrons resembles that of $^{252}$Cf, the detection efficiency should be ~22±2%. We can estimate the actual neutron multiplicities of the suspected anomalies using this value. The results are listed in Table 1. If future data and analysis confirm the anomalies, further investigation will be necessary, including enhancements to the MC codes.

*Table 1. Parameters of Gaussian peaks describing event excess registered in the Anticoincidence (vetoed) spectrum at 583 m.w.e. and smoothed with $\sigma = \sqrt{m}$.*

| | Peak 0 | Peak 1 | Peak 2 | Peak 3 |
|---|---|---|---|---|
| Position ($m$) | 16.2 | 23.3 | 31.5 | 47.1 |
| Sigma | 4.2 | 5.0 | 6.0 | 7.2 |
| $\sqrt{m}$ | 4.0 | 4.8 | 5.6 | 6.9 |
| Area (events) | 16.1 | 11.4 | 5.1 | 6.0 |

| Fraction of the total excess (%) | ~41.8% | ~29.4% | ~13.3% | ~15.4% |
|---|---|---|---|---|
| Actual neutron multiplicity (*M*) | 74 ± 7 | 106 ± 11 | 143 ± 14 | 214 ± 21 |

# 5 NCBJ Setup and Measurements

NCBJ is a Polish acronym for the National Centre for Nuclear Research – one of the largest scientific institutes in Central Europe. This name was chosen for the setup because it was first used at the now-decommissioned underground site of the NCBJ Cosmic Ray Laboratory in Łódź, Poland. The main element of the setup, schematically depicted in Fig. 19, was an array of fourteen proportional counters [19], filled with He-3 at 4 atmospheres and surrounded by a PE moderator forming a 75 × 6.4 × 55 cm$^3$ box of an approximate weight of 25 kg. The helium counters, produced by ZdAJ, Poland [20], are 2.54 cm (1") in diameter and 50 cm in length. Previously, they have been used for neutron background measurements in several European underground laboratories [21, 22, 23] as part of the ILIAS and BSUIN projects [24].

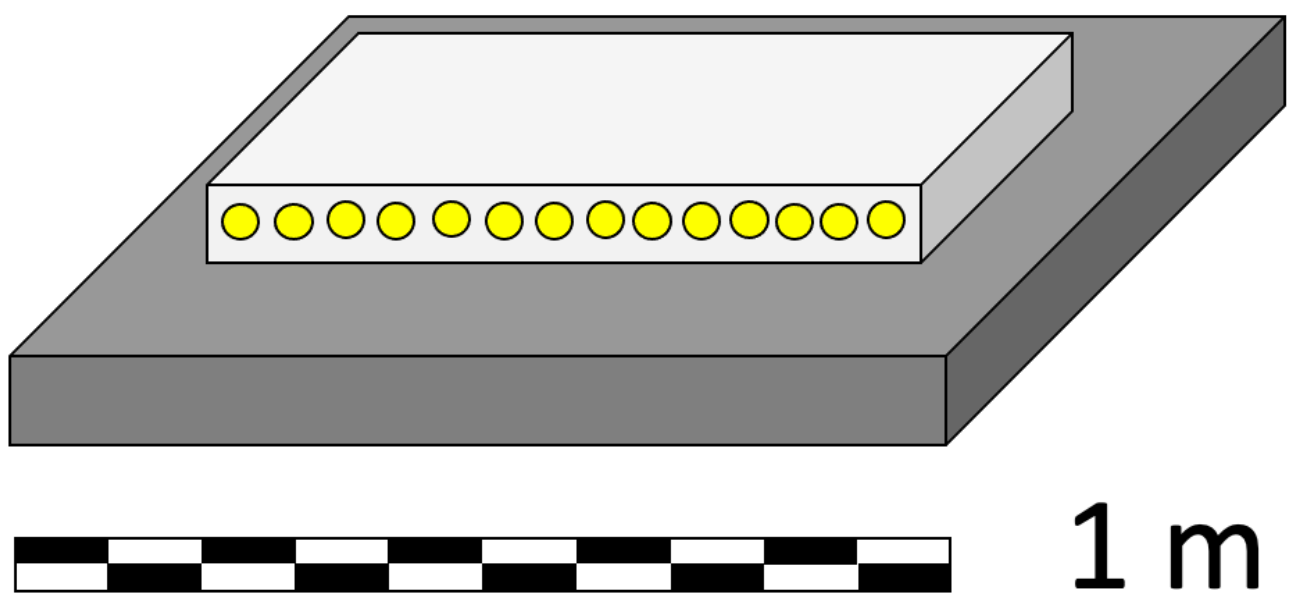

*Figure 19. Schematic diagram of the NCBJ setup. Yellow cylinders, 0.5 m long, 2.54 cm (1") diameter, in a white PE casting represent active volumes of fourteen helium-3 neutron counters. They were placed directly on a 5 cm thick 1 × 1 m$^2$ Pb target.*

The setup was equipped with a custom-made Neutron Data AcQuisition (NDAQ) system [25]. Each detector had a built-in preamplifier and a digitiser sampling the signal for 0.3 ms before and 1.7 ms after the trigger. The trigger was generated whenever any detector registered a valid signal. The sampling time width was sufficient to account for variations in neutron thermalisation. The sampling step and the software allow the detection of multiple neutrons in the same counter if they are spaced by more than 2 – 3 µs. Data-taking was controlled remotely. The extraction of neutron multiplicity from the digitised waveforms is described in [26]. NDAQ is considerably more advanced than the NMDS data acquisition system. At the same time, the number of neutron detectors (14 vs. 60) and, consequently, the detection efficiency (~8% vs. ~22%) are considerably smaller. Also, no in-depth MC simulations of the NCBJ performance and spectra were completed. However, we have collected a background neutron multiplicity spectrum (without a target) and one reference spectrum using a copper (Cu) target in addition to the Pb target.

## 5.1 Measurements at 40 and 210 m.w.e.

The measurement at the NCBJ Cosmic Ray Laboratory in Łódź, Poland, lasted 264.4 hours. The lab was ~13 m underground, corresponding to an overburden of 40 m.w.e. The spectra, recorded in 2010, were first presented in [27] and are shown in Fig. 20 as the lower data series (dark blue points). The experiment aimed to determine the role of particle cascades in generating large neutron multiplicities. Indeed, the largest recorded multiplicity was $m$=45. It was proposed that the most likely mechanism of neutron generation is the interaction of large particle cascades in lead. However, the experiment was intended as a pilot measurement only. Moving on would require a reliable muon tracking system working in delayed coincidence mode with neutron detection. For instance, by integrating neutron detectors with the EMMA experiment [28], located at 210 m.w.e. in the Pyhäsalmi mine [29] in Finland.

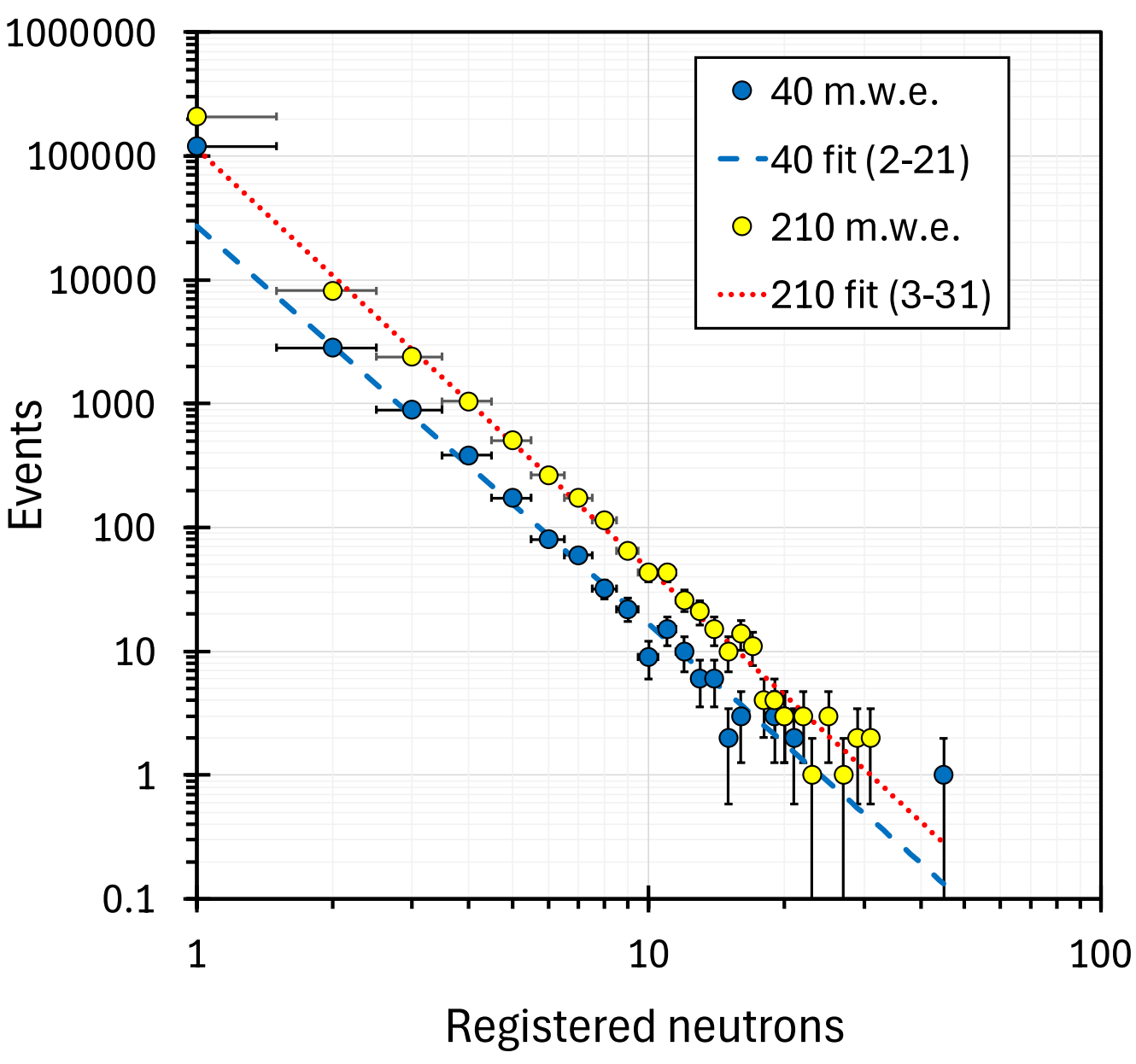

*Figure 20. Neutron multiplicity spectra obtained with the NCBJ setup at 40 (dark-blue points) and 210 m.w.e. (yellow points). The straight lines represent power law fits to the data. The fitting range is indicated in the legend.*

The NCBJ setup was finally relocated to the Pyhäsalmi mine in November 2019. Three measurements were completed at the 210 m.w.e. level before the unavoidable decommissioning of that underground site in November 2022: 344.5 days run with a 565 kg Pb target (a 6.511 metric ton-month exposure), 200.4 days run without a target, and 259.1 days run with a 448 kg Cu target (a 3.870 metric ton-month exposure). Unfortunately, strict COVID-19 mobility restrictions, the defunding of the EMMA project, and the closing of the NCBJ Cosmic Ray Laboratory prevented us from incorporating the muon tracking into neutron measurements. Therefore, the presented analysis relies exclusively on neutron data. The Pb neutron multiplicity spectrum collected at 210 m.w.e. is shown in Fig. 20 (yellow points).

# 6 Evaluation of the NCBJ Spectra

Analysis procedures developed for NMDS data were also applied to NCBJ spectra. Fig. 21 is the NCBJ analogue of Fig. 10. It shows the integrated 210 and 40 m.w.e. data points linearised by the multiplication by $m^{3.5}$ and normalised at $m$=2. As the error bars for both data series are similar, for clarity, in Fig. 21, they were plotted only for the 210 m.w.e data points.

The NMDS analysis (Fig. 10) indicated a possible excess for multiplicities between the normalisation point and $m$~50. When comparing spectra obtained with different efficiency setups, the multiplicity scale must be corrected accordingly. If the NCBJ setup detected the same phenomenon and, accounting for the 2¾ times smaller neutron detection efficiency, there should be a separation between the 210 and 40 m.w.e. data points for multiplicities ranging from the normalisation point up to $m$=18. It is indeed the case. Furthermore, as suggested by the data shown in Fig. 10, the points from a deeper location bulge more in Fig. 21.

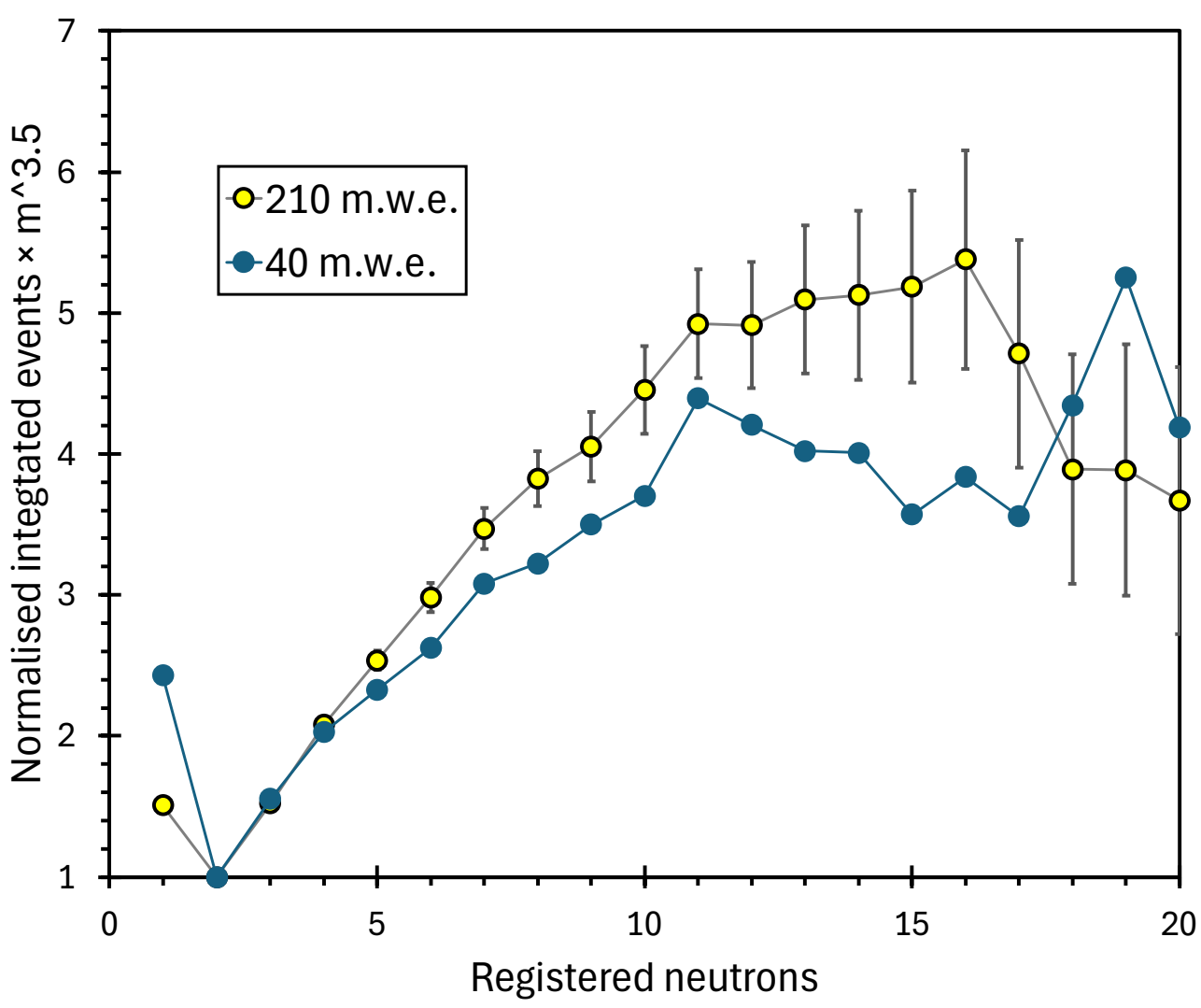

*Figure 21. Integrated NCBJ events at 210 (yellow points with error bars) and 40 m.w.e. (blue points) multiplied by $m^{3.5}$ and normalised to unity at m=2. The analogue representation of the NMDS data is shown in Fig.10.*

## 6.1 Pb neutron spectrum at 40 m.w.e.

The 40 m.w.e. measurement at the NCBJ Cosmic Ray Laboratory in Łódź was the shortest of all experiments evaluated in this paper. Although it lasted only 11 days, it provided a valuable reference for the 210 m.w.e. spectra. Nevertheless, the data shown in Fig. 20 was insufficient for a conclusive outcome. The integrated, background-subtracted spectrum yielded 7.6±21 excess events, i.e. 37±101 events per metric ton-month.

## 6.2 Pb neutron spectrum at 210 m.w.e.

As already mentioned above, no in-depth simulation results exist for the NCBJ setup at 210 m.w.e. However, one of the conclusions of Purdue MC studies [7] was that a power-law fit adequately approximates the muon-induced neutron multiplicity spectrum. Therefore, one can attempt to background-subtract neutron spectra with such a fit. The fit, shown in Fig. 20, used data points $3 \leq m \leq 31$. The integrated excess above the fitted power-law background is shown in Fig. 22. It reaches a maximum at $m$=4, but at lower multiplicities, it fluctuates between large negative and positive numbers. Therefore, we terminate integration at $m$=5 with 75±36 events to avoid possible contamination at $m$=4. These 75±36 events correspond to 12±6 events per metric ton-month.

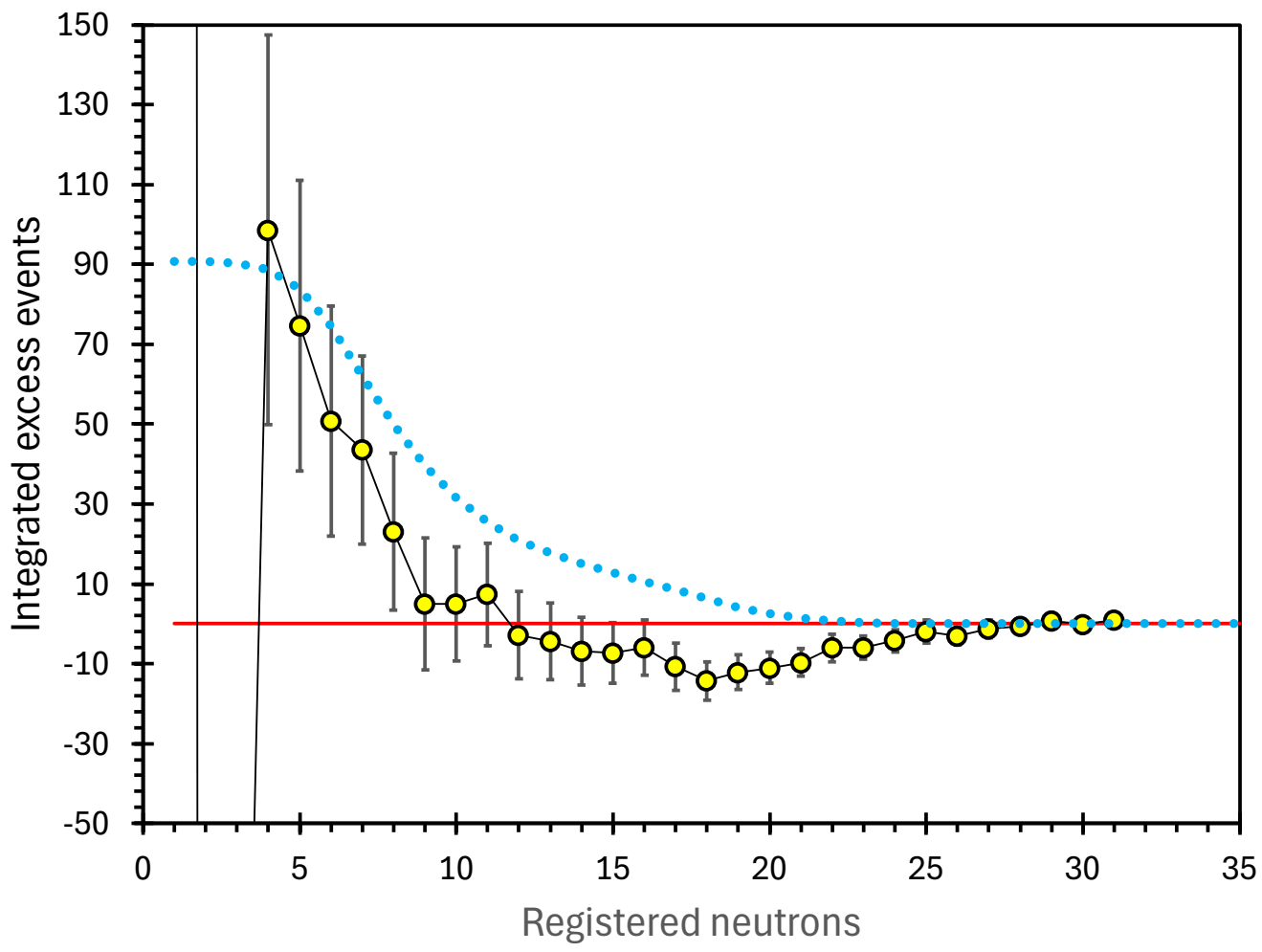

*Figure 22. Integrated excess events registered by the NCBJ setup at 210 m.w.e. The blue dotted line is the prediction based on the NMDS result from 583 m.w.e.*

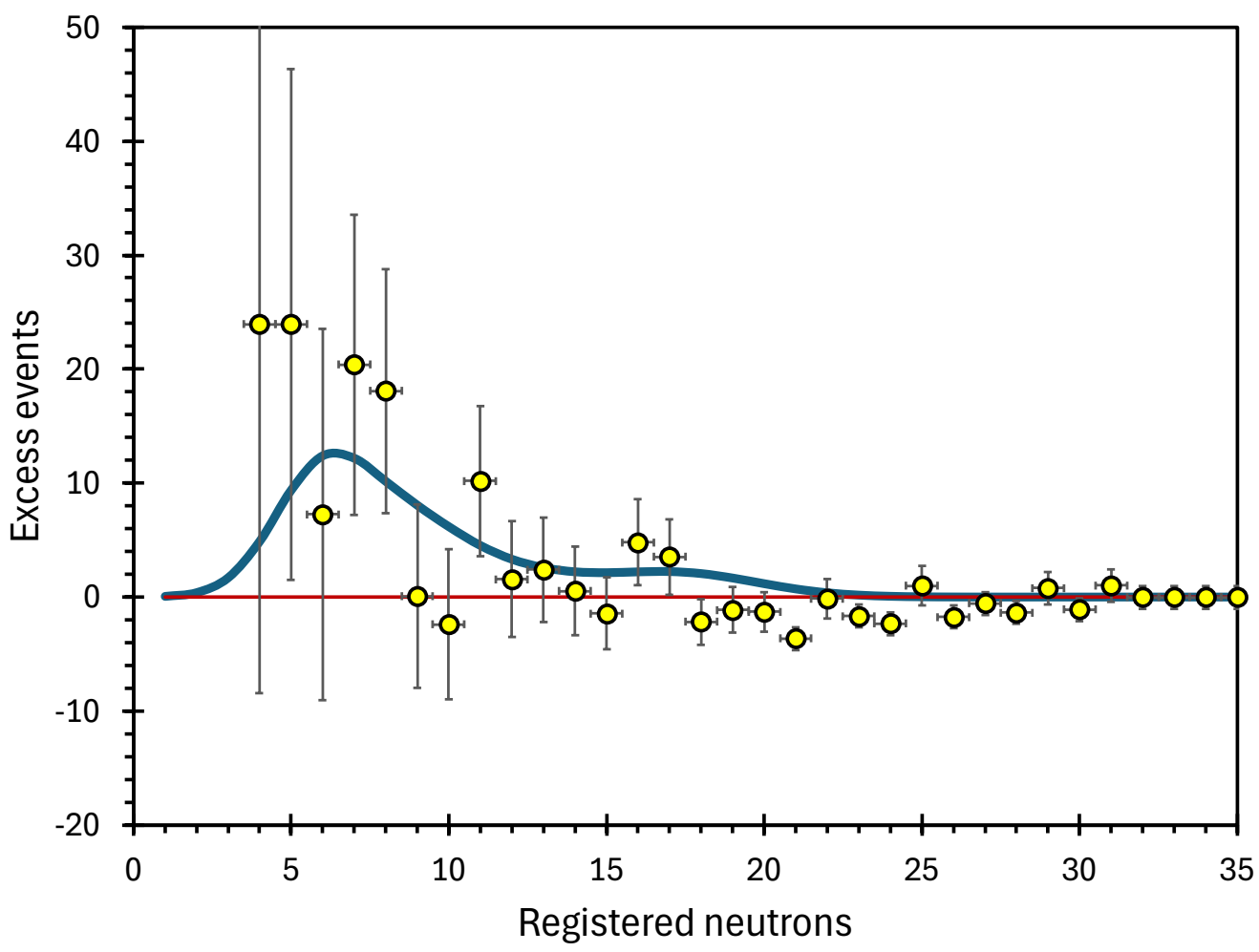

*Figure 23. NCBJ excess events at 210 m.w.e. (points with error bars) superimposed on the 4-peak prediction (blue line) deduced from the NMDS result from 583 m.w.e.*

The next step is to evaluate the shape of the excess spectrum. The blue line in Fig. 23 is the four-peak prediction deduced from the NMDS spectrum at 583 m.w.e. and superimposed on the NCBJ excess events at 210 m.w.e. (points with error bars). The position and sigma parameters of the Gaussian peaks taken from Table 1 were reduced 2¾ times to adjust for the difference in neutron detection efficiencies. The areas from Tab.1 were multiplied by 2.36 – the exposure ratio of 6.5107 ton-month vs. 2.7642 ton-month. While the data are insufficient to confirm shape overlap, there is an apparent similarity between the two. It is also visible in Fig. 22, where the shape of the integrated peaks (blue dotted line) is compared with the integrated excess events at 210 m.w.e. (points with error bars).

## 6.3 Cu neutron spectrum at 210 m.w.e.

The measurements with the NCBJ setup and a 448 kg Cu target took place between August 2021 and November 2022. The total exposure was 3.870 metric ton-months. Fifty standard-size (20 × 10 × 5 cm$^3$) Cu bricks were replaced by fifty Pb bricks. Otherwise, the setup's configuration was not changed. The same analysis procedures were applied to the Cu neutron multiplicity spectrum as previously to the Pb spectrum. The resulting excess spectrum is shown in Fig. 24.

Unlike in the Pb case, no excess was detected with the Cu target. Such an outcome is not obvious, although differences in density, atomic and mass numbers between lead and copper are noticeable. Cu has 8.96 g/cm$^3$ density, Z=29, A= 63.546, while Pb has 11.34 g/cm$^3$, Z=82, and A=207.2. In addition, while Pb is nearly transparent to thermal neutrons, this is not the case with Cu. Longer exposures are needed to search for anomalies with Cu targets.

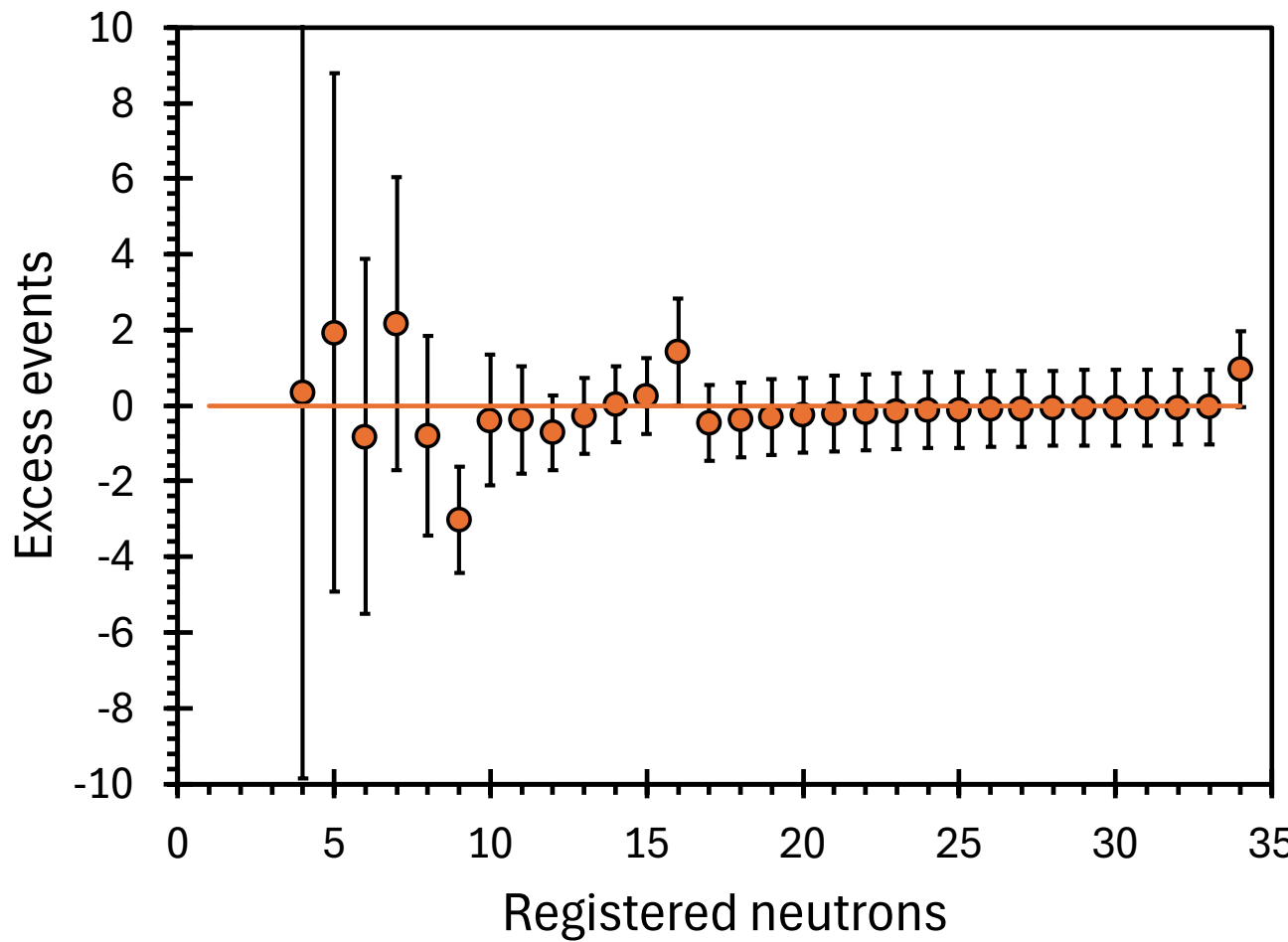

*Figure 24. The excess over the muon-induced neutron multiplicity events from Cu registered at 210 m.w.e.*

# 7 JYFL Setup and Measurements

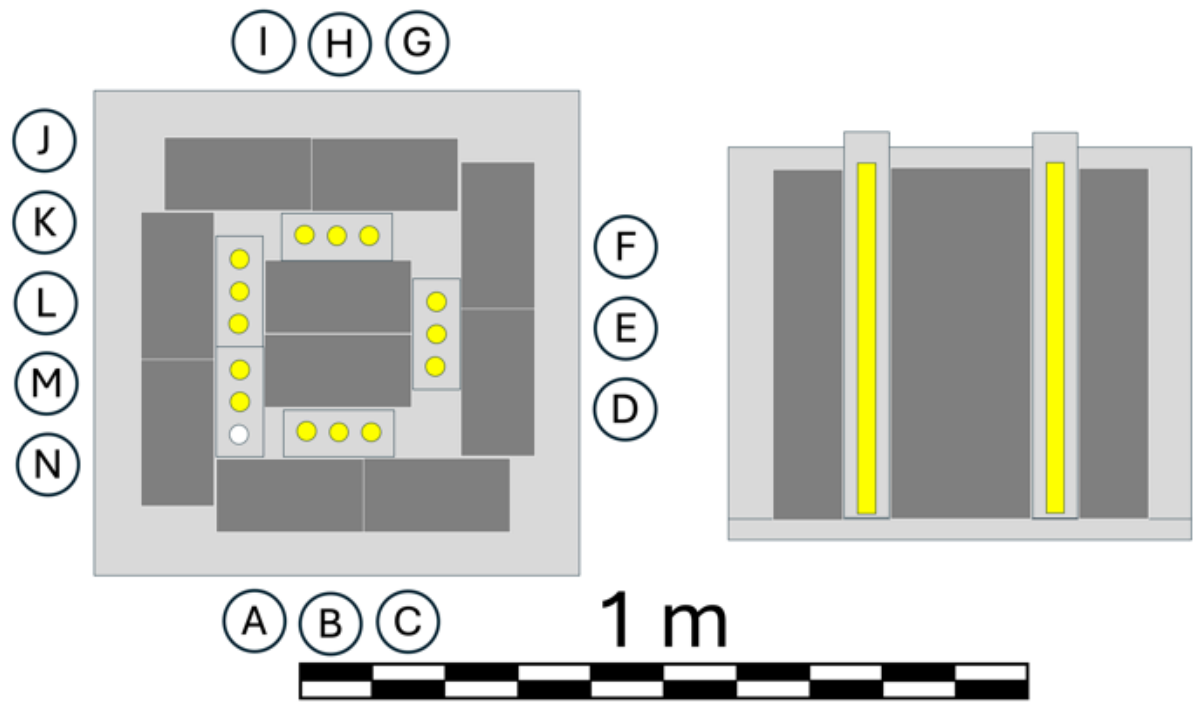

*Figure 25. Horizontal (left) and vertical (right) cross-section of the JYFL setup. It is a 66 × 66 × 56 cm$^3$ PE (light grey) box with 10 piles of ten Pb bricks each (dark grey) instrumented with 14 helium-3 counters (counter's active volume marked yellow). The labelling convention of the detectors is also shown. One detector slot is empty (white). In this compact configuration, there were 10 stacks of 10 bricks, totalling 1154 kg of Pb. The detection efficiency, verified with a $^{252}$Cf source, for neutrons emitted from the two central stacks was 17% and 5-10 % in the eight peripheral stacks.*

To reach the Five Sigma discovery level, we intended to conduct the measurements 1.4 km underground at ~4000 m.w.e. The muon rate at that depth is only 9.5 muons per square meter per day [17]. Consequently, the muon-induced neutron multiplicity spectrum would be highly suppressed, enabling nearly background-free registration of anomalies. The planned setup was described in [30]. As the available funding to realise the proposed experiment was inadequate, we have implemented a significantly simplified version, reusing the 14 NCBJ counters with NDAQ and arranging them as shown in Fig. 25. As the simplified experiment was designed and pre-assembled at the Department of Physics, University of Jyväskylä, known by its Finnish acronym JYFL, we refer to it as JYFL-setup.

During the first 123 days of the experiment, high-multiplicity neutron events occurred at an average monthly rate of 1.5±0.6. In total, six such events were recorded. The event spacing followed the exponential distribution predicted by the Poisson distribution. The detection date, time and time difference between the registered neutrons of each event are listed in Table 2, and the final neutron multiplicity spectrum is shown in Fig. 26. The event with multiplicity $m$=13 is consistent with emission from the central stacks, indicating the actual multiplicity of $M$ ~75±20. The three $m$=6 and two $m$=4 events are consistent with emissions from the peripheral stacks, indicating $M$~76±40. These results are consistent with the most prominent peak (at $M$=74 ± 7) observed at 583 m.w.e. and listed in Tab.1.

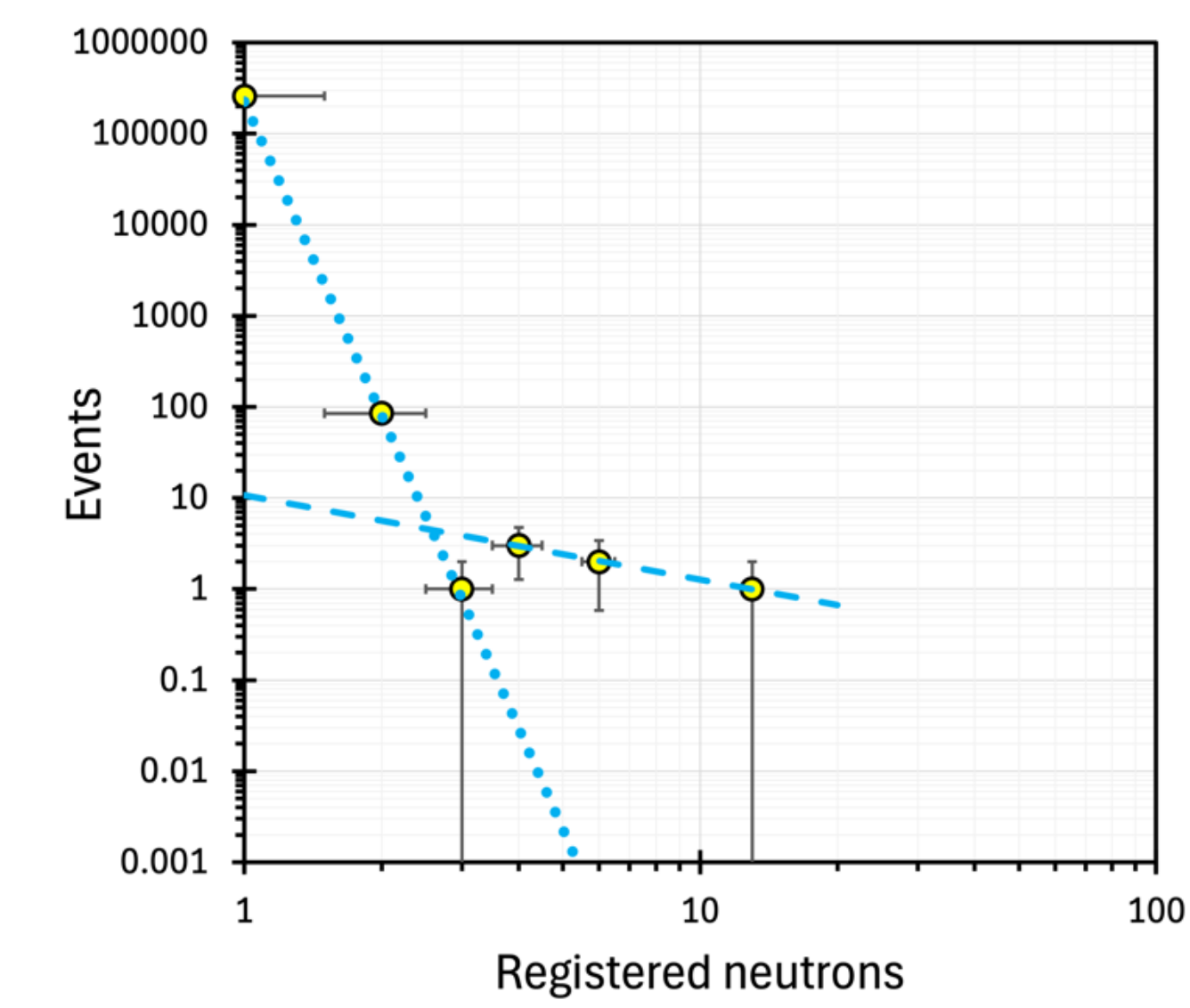

*Figure 26. Neutron multiplicity spectrum measured at 4000 m.w.e. The dotted blue line is a power-law fit to the low-multiplicity part of the spectrum (m<4). The dashed line is the fit to the high-multiplicity part (m>3).*

Strangely, no high-multiplicity events were recorded during the following 542 days of the run. If the rate of 1.5±0.6 events per month (1.3±0.6 per ton-month) determined during the first 123 days of the experiment is correct, the probability of not getting another such event in 18 months is negligible ($<10^{-13}$). Since the detectors worked properly, the most likely cause for the lack of high-multiplicity events after the first 123 days of running is a malfunction of NDAQ. Evidently, the equipment must be inspected and checked, and the measurement must be repeated. Unfortunately, the 1.4 km deep level of the Pyhäsalmi mine is now closed for research projects, and we must rely solely on the available data. Assuming a malfunction soon after the 123$^{rd}$ day of running, we could only accept the results from the first four months of operation. However, it is also possible that the problems with the neutron detectors and the acquisition system started during the relocation from 210 m.w.e. and reinstallation at 4000 m.w.e. In this case, the data from the first four months of measurement were also unreliable. Only a new measurement may settle these questions.

# 8 Discussion

The main problems in anomaly searches are limited statistics and inadequate knowledge of muon-induced neutron spectra. The former may be solved with better experiments running longer and using larger targets; the latter is more challenging. While a generalised assumption that a power-law function describes muon-induced spectra is justified, the actual spectra show a more complex picture. The example from the shallow site is shown in Fig. 7, where the muon-suppressed (vetoed) NMDS data at 3 m.w.e. is displayed in the log-log scale and fitted with a power-law dependence in the neutron multiplicity range $3 \leq m \leq 20$ (solid orange line, $p$=4.21) and $20 \leq m \leq 70$ range (dashed blue line, p=0.318). There are two very different trends. If the red line represents muon-induced neutron production in Pb, the physics behind the dashed blue trend line needs to be investigated.

*Table 2. Parameters of all high-multiplicity neutron events collected with the JYFL setup. The trigger time of each event is given with one-minute precision. The trigger starts the high-accuracy clock, so all the subsequent signals are timed with a one-microsecond step. The triggering pulse has the time after trigger equal to zero.*

| **Multiplicity** | **Date** | **Trigger time** | **Comment** |
|---|---|---|---|
| 6 | 6.12.2022 | 21:04 | BOT |
| **Detector #** | **Placement** | **Time after trigger [us]** | |
| 2018-02 | B | 0 | neutron |
| 2018-04 | D | 1 | neutron |
| 2019-03 | M | 1 | neutron |
| 2018-03 | C | 3 | neutron |
| 2018-01 | A | 56 | neutron |
| 2018-06 | F | 56 | neutron |

| **Multiplicity** | **Date** | **Trigger time** | **Comment** |
|---|---|---|---|
| 6 | 10.12.2022 | 3:18 | BOT |
| **Detector #** | **Placement** | **Time after trigger [us]** | |
| 2018-05 | E | 0 | neutron |
| 2018-04 | D | 13 | neutron |
| 2018-03 | C | 20 | neutron |
| 2018-01 | A | 42 | neutron |
| 2018-01 | A | 92 | neutron |
| 2018-03 | C | 229 | neutron |

| **Multiplicity** | **Date** | **Trigger time** | **Comment** |
|---|---|---|---|
| 6 | 15.1.2023 | 14:16 | LEFT |
| **Detector #** | **Placement** | **Time after trigger [us]** | |
| 2018-05 | E | 0 | neutron |
| 2018-01 | A | 8 | neutron |
| 2019-01 | L | 30 | neutron |
| 2018-09 | I | 138 | neutron |
| 2019-01 | L | 202 | neutron |
| 2018-01 | A | 469 | neutron |

| **Multiplicity** | **Date** | **Trigger time** | **Comment** |
|---|---|---|---|
| 4 | 21.1.2023 | 14:19 | LEFT |
| **Detector #** | **Placement** | **Time after trigger [us]** | |
| 2018-10 | J | 0 | not-neutron |
| 2019-01 | L | 59 | neutron |
| 2018-09 | I | 70 | neutron |
| 2018-01 | A | 95 | neutron |
| 2019-09 | K | 117 | neutron |
| 2018-08 | H | 435 | not-neutron |

| **Multiplicity** | **Date** | **Trigger time** | **Comment** |
|---|---|---|---|
| 13 | 20.3.2023 | 8:46 | CENT |
| **Detector #** | **Placement** | **Time after trigger [us]** | |
| 2018-06 | F | 0 | overflow |
| 2018-09 | I | 4 | neutron |
| 2019-01 | L | 4 | neutron |
| 2018-07 | G | 5 | neutron |
| 2018-08 | H | 6 | neutron |
| 2018-05 | E | 19 | neutron |
| 2018-02 | B | 30 | neutron |
| 2019-09 | K | 36 | neutron |
| 2018-08 | H | 42 | neutron |
| 2019-03 | M | 43 | neutron |
| 2018-06 | F | 55 | neutron |
| 2018-05 | E | 71 | neutron |
| 2018-09 | I | 106 | neutron |
| 2018-09 | I | 216 | neutron |

| **Multiplicity** | **Date** | **Trigger time** | **Comment** |
|---|---|---|---|
| 4 | 28.3.2023 | 17:11 | ? |
| **Detector #** | **Placement** | **Time after trigger [us]** | |
| 2018-09 | I | 0 | neutron |
| 2018-01 | A | 7 | neutron |
| 2018-01 | A | 34 | neutron |
| 2018-08 | H | 226 | neutron |

At greater depths, our data are scarce, but the crossover point appears to shift to noticeably lower multiplicities. If our interpretation is correct, it is $m \approx 20$ at 3 m.w.e. (Fig. 7), and $m \approx 5$ at 1166 m.w.e. (Fig. 3). The dotted blue line in Fig. 3 is the power-law fit of the data at $2 \leq m \leq 5$ ($p$=7.26), and the dashed blue line at $5 \leq m \leq 32$ ($p$=0.623). The solid red line is the MC prediction reduced by a factor of two, as explained in section 4.3. The red dashed line is the Purdue fit. This one-component fit and the MC prediction overlap with only three points in the mid-section. The mismatch between the MC and 1166 m.w.e. data is also visible in Fig. 14, where linearised and normalised spectra are compared with the calculated MC trends. Unlike at 583 m.w.e., MC does not describe the 1166 m.w.e. neutron multiplicity spectrum. Finally, a similar two-component spectrum appears in the JYFL 4000 m.w.e. data (Fig. 26).

The NMDS 583 m.w.e. spectrum in Fig. 11 also reveals an interesting feature. While most of the data points follow the MC prediction (red line) and the $m$=1 point is evidently affected by the ambient neutron background, the points $2 \leq m \leq 6$ follow their own trend (thin green dashed-dotted line, $p$=4.21) that is practically identical to that for total (thin dashed blue, $p$=4.20) and vetoed (thin purple dotted line, $p$=4.20) NMDS 3 m.w.e. trends.

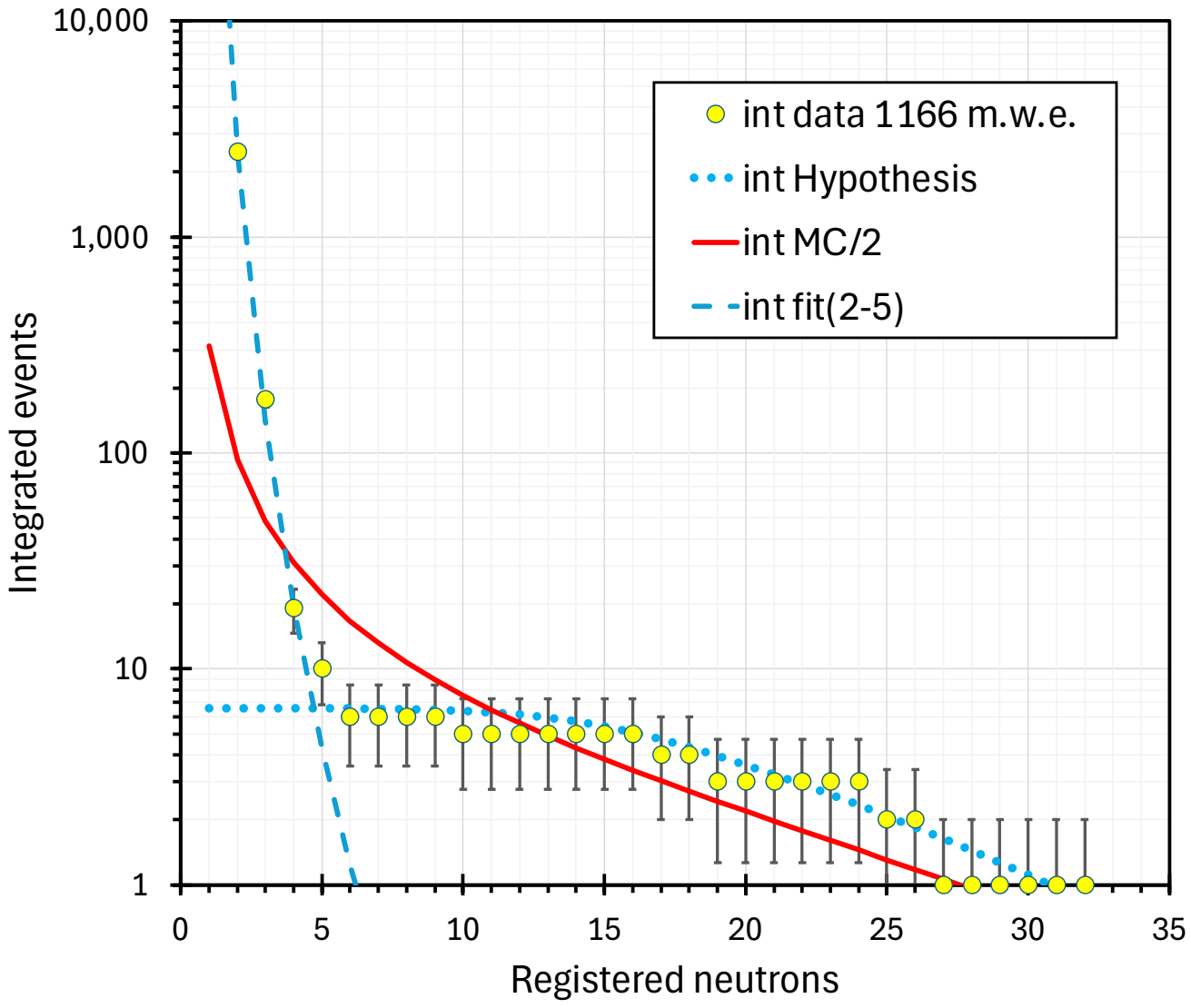

*Figure 27. Integrated neutron multiplicity measured at 1166 m.w.e. The solid red line is the MC prediction. The blue dotted line is the contribution from the assumed anomalies, based on the NMDS result from 583 m.w.e. The red line is the MC simulation reduced by a factor of two, as explained in section 4.3. The blue dashed line is a power-law fit to the data with multiplicity* $2 \leq m \leq 5$.

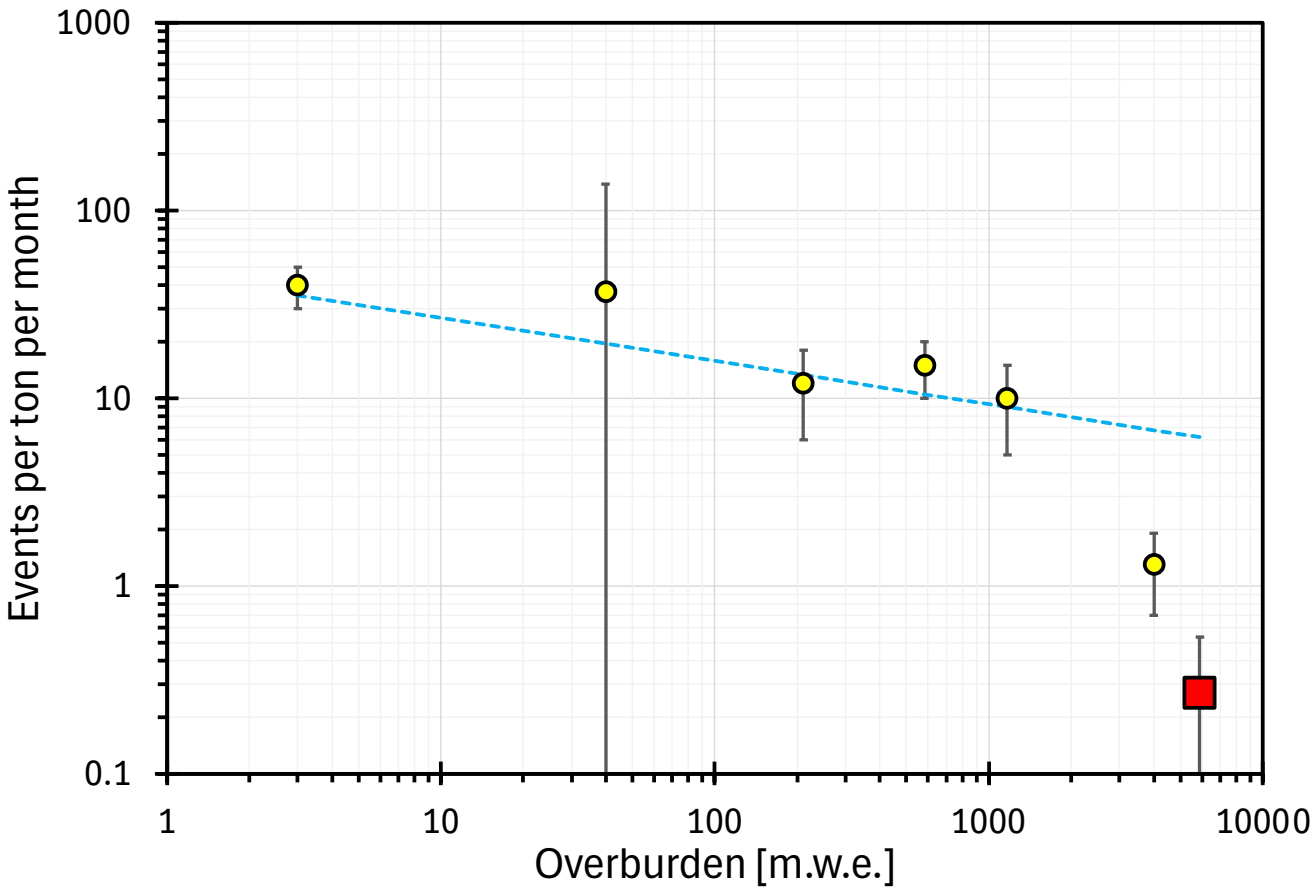

*Figure 28. Anomalous events per ton per month as a function of overburden. Error bars reflect statistical uncertainties only. The dashed trend line is based on data from locations with an overburden of less than 1200 m.w.e. The red square is based on the preliminary HALO spectrum presented at TAUP 2023.*

Evidently, the simplified one-power-law description of the neutron multiplicity spectra becomes insufficient already for the surface data collected with muon suppression (Fig. 7).

Therefore, we treat the second component as an anomaly not reproduced by MC simulations. Since the same problem occurs in the data collected and at overburdens exceeding ~1000 m.w.e. (Fig. 3 and 26), we assumed in the analysis of the data from 1166 and 4000 m.w.e. that the muon-induced contribution is described by the first (steeper) component and used it for background subtraction. With this assumption, unlike MC prediction (red line), the 1166 m.w.e. spectrum is consistent with the four-peak anomaly hypothesis (the blue dotted line in Fig. 27). After integration, the anomaly hypothesis yields 5.8±3.2 events, i.e., 10±5 events per metric ton-month.

The four-month measurement with the JYFL setup at 4000 m.w.e. detected six high-multiplicity anomalies separated from the steep dependence marked as a blue dotted line in Fig. 26 and interpreted by us as a muon-induced portion of the spectrum. The outcome of the measurements is summarised in Table 3 and shown as a function of overburden in Fig. 28. The data originate from measurements at five locations, ranging in depth from 3 to 4000 m.w.e. The yield of excess neutrons emitted from Pb gently decreases with depth. The dashed trend line in Fig. 28 is based on the five shallowest points. This shallow trend predicts 6.7 excess events per metric ton per month at 4000 m.w.e., while the measured value is 1.3 ± 0.6. As explained above, the lower-than-expected outcome may be related to NDAQ malfunctions arising from the hasty relocation of the setup from the 210 to the 4000 m.w.e. site. On the other hand, when the anomalies are plotted as a function of muon flux (Fig. 30), all data points align reasonably well.

Nevertheless, at this point, we cannot be certain of the anomalies' reality. For that, we need good quality data from 1000 m.w.e. or deeper. Also, we do not have a solid explanation of the nature and properties of the anomalies. Nevertheless, high-multiplicity neutron sources other than cosmic-ray-induced muons remain a valid hypothesis, as the event rate per ton at different depths does not correlate directly with the muon flux. For instance, the excess may be related to an indirect WIMP decay/annihilation via weak interaction with a Pb nucleus. If such an indirect process occurs, the setup consisting of a Pb target surrounded by neutron counters would be able to detect the indirect Spin Independent WIMP-nucleon annihilation cross section at the level of ~$10^{-45}$ cm$^2$ for WIMP masses between 0.1 and 10 GeV/c$^2$. For indirect Spin-Dependent annihilation, the sensitivity limit would be ~$10^{-42}$ cm$^2$. These cross-section limits were recently recalculated by the Purdue group [31]. Before further speculations, we propose confirming the existence and properties of the suspected anomalies in neutron multiplicity spectra with a dedicated experiment outlined in the Outlook section.

*Table 3. Observed anomalies and calculated muon fluxes.*

| **Overburden** m.w.e. | **Muon flux** $\mu$/m$^2$/s | **Anomaly** | **Stat. Error** |
|---|---|---|---|
| | | events/ton/month | |
| 3 | 1.80E+02 | 40 | 10 |
| 40 | 5.04E+01 | 37 | 101 |
| 210 | 1.19E+00 | 12 | 6 |
| 583 | 9.74E-02 | 15 | 5 |
| 1166 | 1.39E-02 | 10 | 5 |
| 4000 | 1.12E-04 | 1.3 | 0.6 |

## 8.1 Background analysis

Not all neutrons detected by our experimental setups originate from the target. In fact, most of the low-multiplicity neutron events arise from muon-induced neutron emissions in the surrounding rocks, fission, and other sources. However, due to limited acceptance, all high-multiplicity events come from the immediate vicinity of the detectors. For example, if 100 neutrons were emitted from a location 1 m away from the centre of the NMDS setup, only about 3 neutrons would enter the setup, since at 1 m we cover only 3% of the 4π solid angle. Given that the NMDS detection efficiency is ~22%, we would detect at most 1 neutron from such an event (3 x 0.22 = 0.66). To compare, 1 m is roughly the distance from the floor to the setup, while the distances from the ceiling and cavern walls are considerably larger (3-6 m).

The alternative scenario involves a pileup of neutrons from multiple separate muon-induced reactions. However, considering the very low muon flux underground, such pileups are statistically impossible in deep locations. Also, detector contaminations can be ruled out, as there are usually only 3 to 4 neutrons per fission. Neither pileup nor fission can add up to mimic a high-multiplicity event.

In conclusion, only sources within or immediately surrounding the active volume can cause the registration of high-multiplicity neutron events. Since there are no significant amounts of other high-Z materials within the 2 x 2 x 2 $m^3$ cube around the detection setup, such as the NMDS, the only plausible explanation is that the Pb target is the sole source of these high-multiplicity neutron events.

## 8.2 HALO experiment

One ongoing measurement that could provide valuable data for studying anomalies in muon-induced neutron spectra is HALO – the Helium and Lead Observatory [32]. HALO has a 78,624 kg Pb target instrumented by 128 ultra-low-activity He-3 neutron counters, providing 368 m of active detector length. Located at SNOLAB with a 6000 m.w.e. overburden, HALO has been taking data since 2012 and is part of the SuperNova Early Warning System (SNEWS) [33].

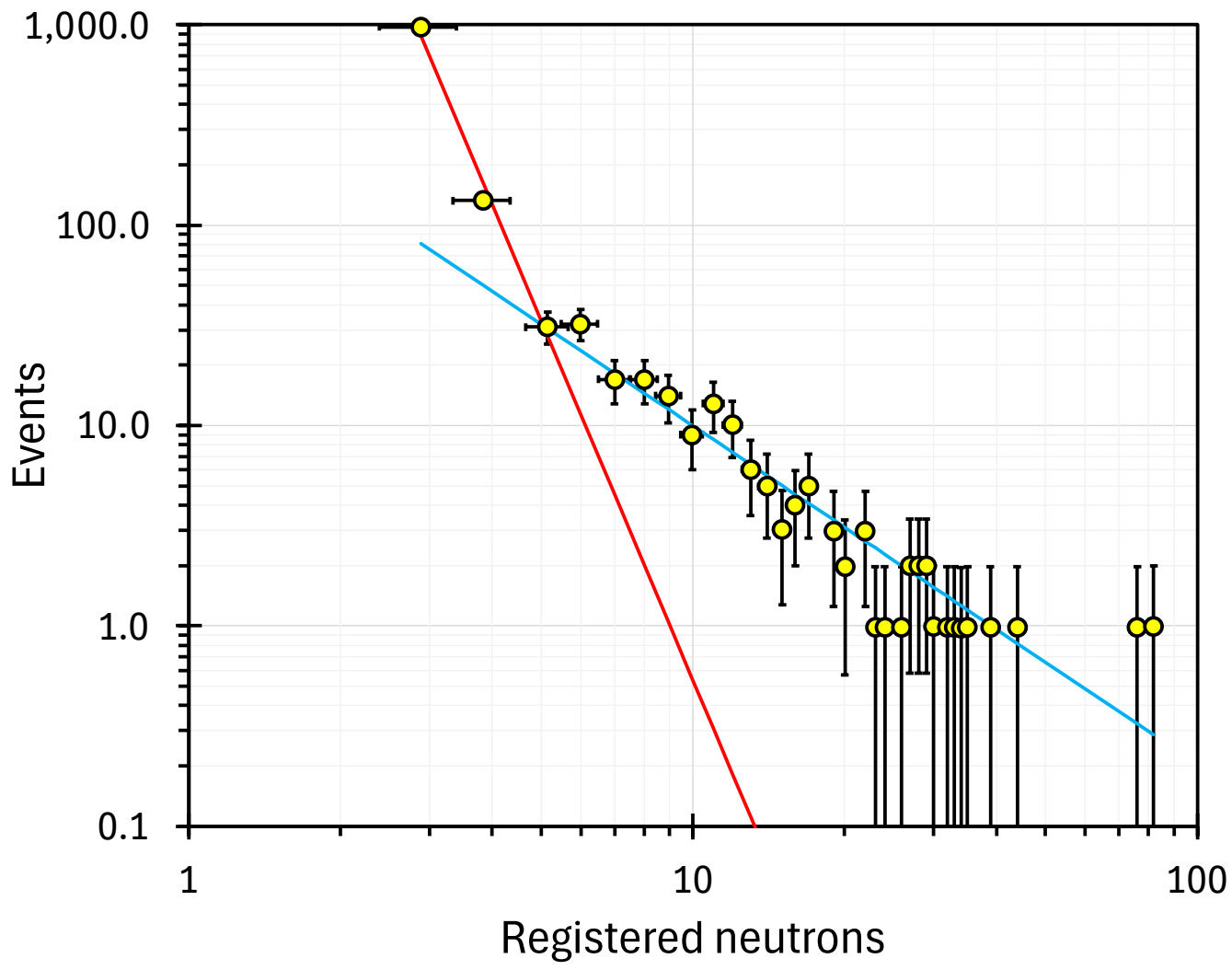

*Figure 29. Preliminary HALO spectrum extracted from TAUP 2023 poster. The red trendline is based on the first 3 data points. The blue trendline fits the remaining points except the two at the end of the spectrum.*

Despite its potential, the analysis, simulation, and interpretation of HALO data are complicated by the detector's irregular 'Swiss cheese'-like geometry, the strong position dependence of neutron detection efficiency, and the lack of position sensitivity in the He-3 counters. In contrast, the setup proposed in the Outlook section, although smaller in scale, is better suited to neutron multiplicity studies. It offers significantly improved instrumentation, position sensitivity and a much higher detector density, with 60 meters of active detector length per metric ton of target material, compared to only 4 m/ton in HALO.

To date, the only publicly accessible HALO neutron multiplicity spectrum was presented at the 2023 TAUP conference in Vienna [34]. The spectrum is marked as 'preliminary' and covers 5.6 years of HALO operation. Fig. 29 shows the HALO spectrum from the TAUP 2023 poster, replotted on the log-log scale, and supplemented with statistical error bars.

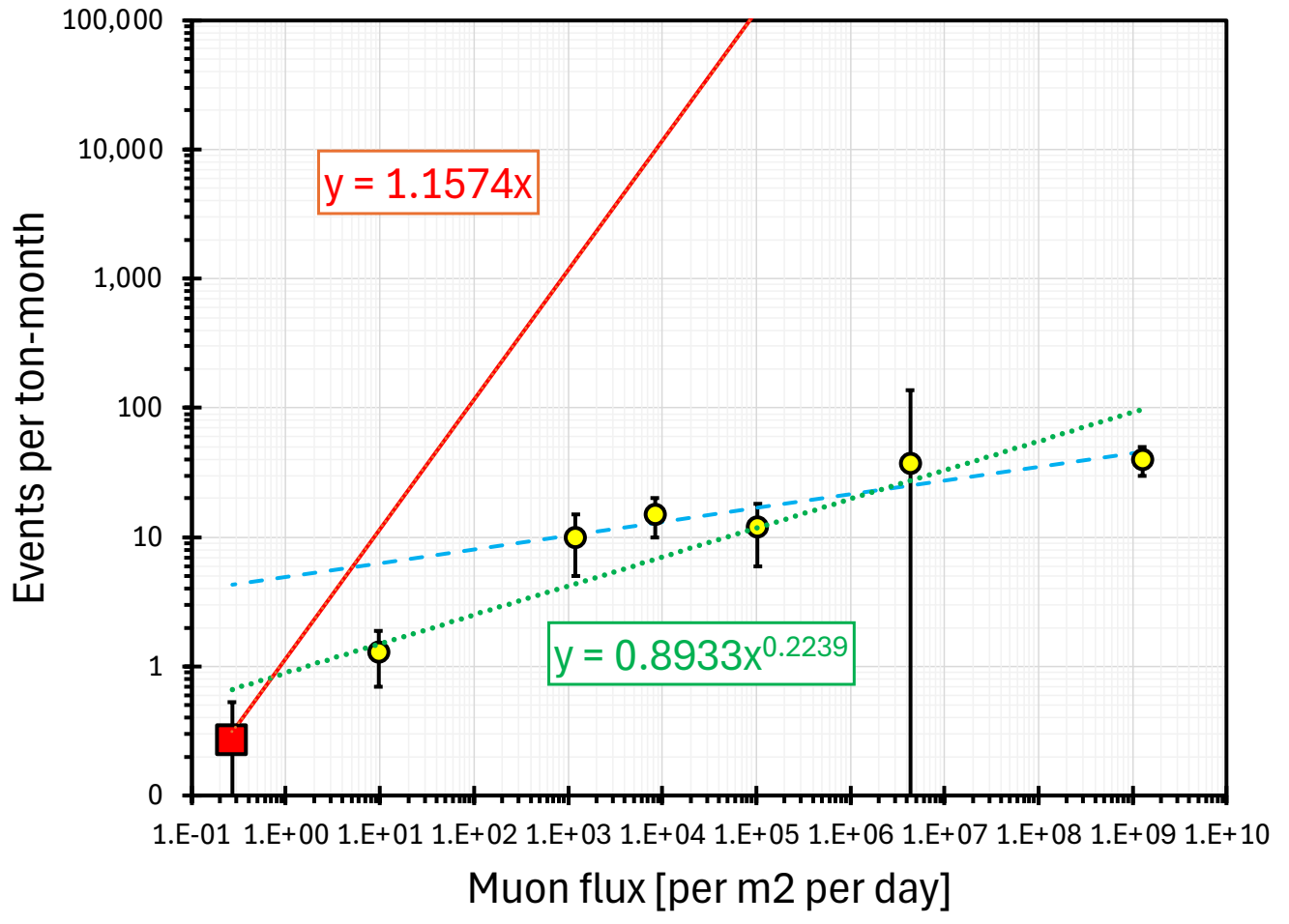

*Figure 30. Data from Fig. 28 replotted as a function of the muon flux. The blue dashed line is the trend from Fig. 28. The green dotted line is the trend based on all data points. The solid red line shows direct dependence on the muon flux normalised to the known muon flux at the HALO site.*

During the 5.6 years (2,054 days) of exposure, only 4,108 muons have reached HALO [34]. And yet, the spectrum in Fig. 29 contains 1,302 events with multiplicities greater than 2. We get over half a million events if we extrapolate the red line in Fig. 29 to the missing multiplicities (m=1 and 2), indicating a substantial background contribution. As clarified in a private communication, the spectrum in Fig. 29 has not been unfolded and may therefore contain confounding effects from neutron efficiency as a function of neutron multiplicity, etc. Nevertheless, the preliminary HALO spectrum shows exactly the characteristics we predicted. Like at 1166 (Fig. 3) and 4000 m.w.e. (Fig. 26), there appear to be two distinct components in the HALO neutron multiplicity spectrum. If we extrapolate and integrate the blue trend line in Fig. 29, we get 1412 events, that is 0.27 events per metric ton per month. If the blue component in Fig. 29 represents anomalies, we can add it (as a red square) to Fig. 28 and 30. As both the HALO spectrum and our analysis of their data are preliminary, we attribute arbitrary 100% error bars to that point (red square). It follows the trend of a mild muon flux dependence set by our results (the green dotted line in Fig. 30). For comparison, the thick red line in Fig. 30 illustrates the trend for events directly proportional to the muon flux.

# 9 Outlook

Although the persistence of the observed anomalies is compelling, their statistical significance falls short of the desired five sigma discovery level. To tackle this challenge, we have established the **NEMESIS** Collaboration, dedicated to **NEutron MEasurementS In Subterranean locations**. Following the closure of the Pyhäsalmi mine, we are seeking new collaborators and funding to relocate the measurements to a new deep underground site and amend the data with extensive MC simulations. Further, in partnership with Proportional Technologies, Inc., we are developing a high-efficiency, low-background, low-cost neutron detector array using boron-coated straws (BCS). Each straw is a copper tube, 15 mm in diameter, lined on the inside with a thin layer of boron carbide ($^{10}B_4C$) that is 96% enriched in the $^{10}B$ isotope. The length of the BCS is adapted to the requirements. Thermal neutrons are converted into secondary charged particles via the B(n,α) reaction: $^{10}B + n \rightarrow {}^{7}Li + \alpha$. BCS-based detectors [35] are employed, e.g., to detect radioactive materials by the United States Department of Homeland Security [36] and by the LOKI broadband Small Angle Neutron Scattering (SANS) instrument at the European Spallation Source.

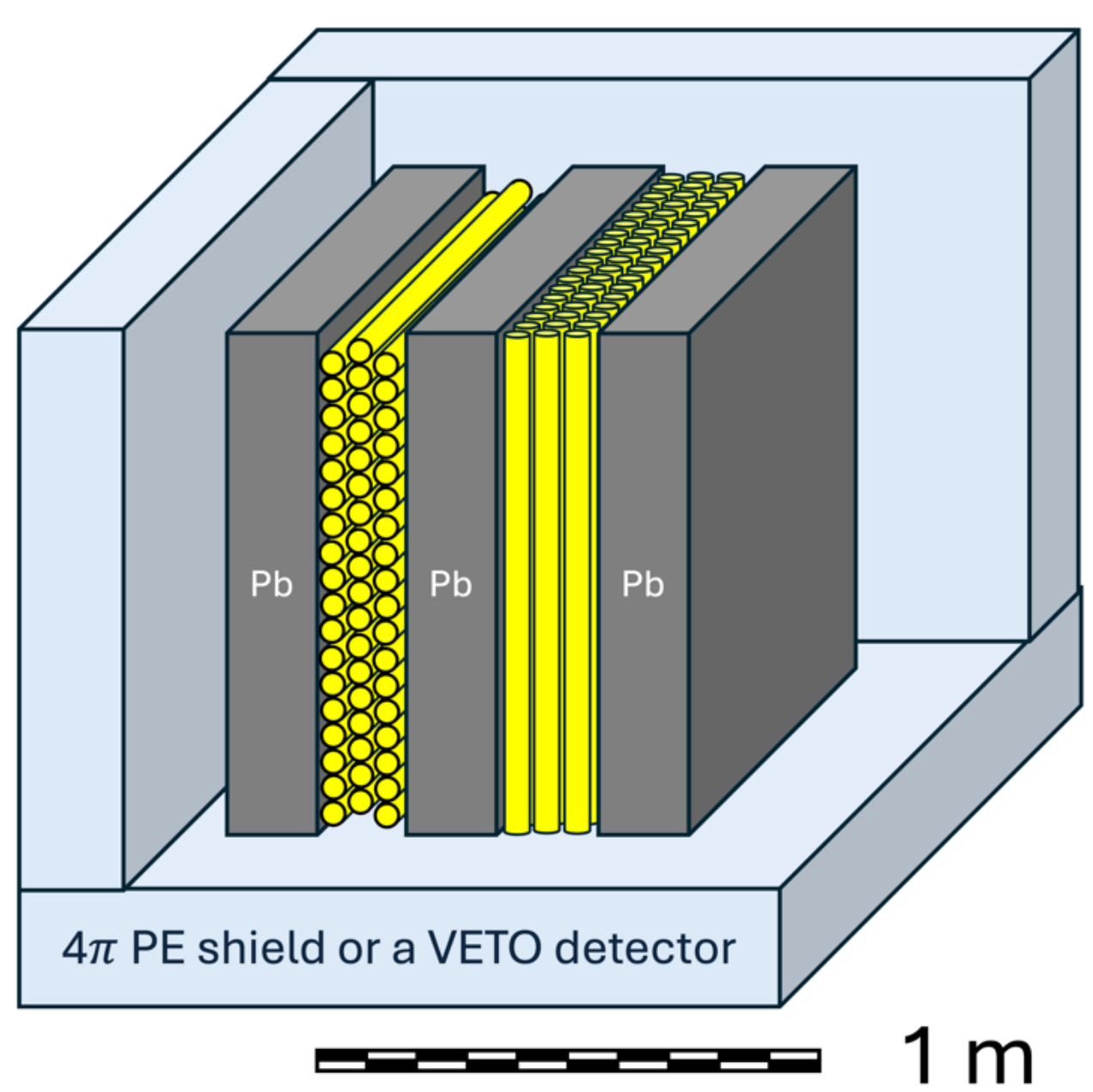

*Figure 31. Concept of a setup for detecting neutron multiplicity spectra emitted from a 3.4-ton Pb target. Each of the three Pb walls is made of 100 standard-size bricks. The 90 boron-coated straw (BCS) tubes in the first LAND detector are aligned vertically, and in the second LAND detector, they are aligned horizontally. The system is surrounded from all six sides (only three are depicted) with a passive (HD PE) or, funding permitting, an active shield.*

The current conceptual design of the proposed NEMESIS setup is illustrated in Fig. 31. It depicts the active elements (BCS) of two position-sensitive Large Area Neutron Detectors (LAND) placed between three walls made of standard Pb bricks. The total Pb mass in this configuration will be 3,402 kg. With hits in multiple straws, we can use the inverse square distance prediction to identify one of the three Pb walls where the emission of a high-multiplicity neutron burst occurred. Each LAND comprises 90 BCS. A digitiser-based data acquisition system records the time and position of the fired BSC for each detected neutron. As the two LANDs will be aligned perpendicularly to each other, the setup will ascertain the XY coordinates of high-multiplicity events, allowing us to determine the topology of each neutron burst. If emitted by a hypothetical WIMP decay/annihilation, all neutrons in a high-multiplicity event should originate from a point-like source in the Pb target. A spallation-like mechanism would generate neutrons along the muon trajectory, while multiple muon-induced hadronic reactions would indicate several point-like origins. The new experiment should distinguish between these scenarios. During the first year of operation, the setup in Fig. 31 would register an order of magnitude more high-multiplicity events than all our previous measurements combined. The setup can easily be expanded if needed by adding, at a moderate cost, alternating target/LAND layers. With additional funding, BSDs can be embedded directly in the Pb target that would also serve as a moderator.
The final refinement of the setup will be a plastic-scintillator-based charged-particle tracking detector replacing the passive 4π PE shielding surrounding the target and LANDs. The 4π coverage is desired to distinguish between the traversing muons and particles emitted from the target. The former would trigger two sides of the veto cube, and the latter would trigger only one. In addition, potential back-to-back target emissions would have similar registration time at the opposite sides of the scintillator cube, while the traversing muons would be separated by the muons' flight time. Funding permitting, we look forward to realising this project soon.

# 10Summary

- Neutron multiplicity spectra from CR muon interactions underground are routinely approximated by a $k \times m^{-p}$ power-law function, where the slope parameter *p* remains roughly constant for the given configuration, while the amplitude parameter *k* diminishes with the declining muon flux.
- Our measurements with the NMDS setup (Fig. 1 and 6) at 3, 583 and 1166 m.w.e., the NCBJ setup (Fig. 19) at 40 and 210 m.w.e., and the JYFL setup (Fig. 25) at 4000 m.w.e. indicate the probability of an additional anomalous component in neutron multiplicity spectra.
- If confirmed, the existence of the additional component would impact background studies for deep-underground experiments, revise MC simulation codes, and may even deepen our understanding of particle interactions.
- The evidence acquired so far for the anomalies (Figs. 28, 30, and Table 3) falls short of the five-sigma discovery threshold but merits follow-up experiments.
- To cross the five-sigma discovery threshold, an order of magnitude higher statistics are required, which are reachable with the setup described in the Outlook section during the first year of operation.
- If our interpretation is correct, the anomalous component would dominate the high-multiplicity part of the neutron multiplicity spectra at overburdens exceeding 1000 m.w.e..
- The absolute value of the excess appears to drop slightly with overburden (Fig. 28) and increase with muon flux (Fig. 30), but measurements with a rudimentary muon veto used with the NMDS setup at 583 m.w.e. indicate a lack of direct correlation with traversing muons (Fig. 16).
- The smoothed neutron multiplicity spectra at 583 m.w.e. hint at a possible structure (Fig. 17) resembling four Gaussian peaks (Fig. 18). The statistics are insufficient for conclusive assessment, but the peaks would be consistent with anomalous neutron emissions with multiplicity *M*=74 ± 7, *M*=106 ± 11, *M*=143 ± 14, and *M*=214 ± 21. The shapes of the excess events postulated at 210 and 1166 m.w.e. (Figs. 22, 23, and 27) are also consistent with such a hypothesis.

## Acknowledgements

Many scientists initially involved in this project are no longer with us or have left the research field. We acknowledge the contributions of Alexander A. Rimsky-Korsakov, Nikolay A. Kudryashev, and their former team members to the commencement of this research. We express our gratitude to Callio Lab [https://calliolab.com] for providing access to the underground locations and infrastructure of the Pyhäsalmi mine. We gratefully acknowledge financial support from the Helsinki Institute of Physics, TechSource Incorporated, and the Wihuri Foundation. Our special thanks go to the Kerttu Saalasti Institute team at the University of Oulu, including Dr Ossi Kotavaara, Mr Jari Joutsenvaara, and Ms Julia Puputti, for their assistance and support during the setup and conduct of the measurements.

## References

[1] H. Kluck, Measurement of the cosmic-induced neutron yield at the Modane underground laboratory (2015), https://link.springer.com/book/10.1007/978-3-319-18527-9 ; first published (2013) https://publikationen.bibliothek.kit.edu/1000039837

[2] O. Nairat , J. F. Beacom , and S. Weishi Li, Neutron tagging can greatly reduce spallation backgrounds in Super-Kamiokande, PR D 111, 023014 (2025), https://doi.org/10.1103/PhysRevD.111.023014

[3] J. Albrecht et al., The Muon Puzzle in cosmic-ray induced air showers and its connection to the Large Hadron Collider, Astrophysics and Space Science (2022) 367:27, https://doi.org/10.1007/s10509-022-04054-5

[4] D.-M. Mei and A. Hime, Phys. Rev. D 73, 053004 (2006), https://arxiv.org/abs/astro-ph/0512125

[5] V. Pec et al., Muon-induced background in a next-generation dark matter experiment based on liquid xenon, 2023, https://arxiv.org/pdf/2310.16586

[6] M. Palermo, The Muon-Induced Neutron Indirect-Detection EXperiment: MINIDEX, 2016 https://edoc.ub.uni-muenchen.de/19575/1/Palermo_Matteo.pdf

[7] Haichuan Cao and David Koltick, Cosmic Ray Induced Neutron Production in a Lead Target, 2024, https://arxiv.org/pdf/2401.11280 ; see also Haichuan Cao and David Koltick, Dark Matter Induced Neutron Production Search Limits, https://arxiv.org/pdf/2506.22659

[8] T. E. Ward et al., APS April Meeting 2019, https://meetings.aps.org/Meeting/APR19/Session/G17.1 . Accessed: 2025-02-08

[9] W.H. Trzaska et al., New NEMESIS Results, PoS(ICRC2021)514, https://pos.sissa.it/395/514/pdf

[10] W.H. Trzaska et al., 2021, DM-like anomalies in neutron multiplicity spectra, J. Phys.: Conf. Ser. 2156 012029, https://iopscience.iop.org/article/10.1088/1742-6596/2156/1/012029/pdf

[11] W.H. Trzaska et al., New Evidence for DM-like Anomalies in Neutron Multiplicity Spectra, PoS(TAUP2023)083, https://pos.sissa.it/441/083/pdf

[12] J. Billard et al., Direct Detection of Dark Matter – APPEC Committee Report, 2022 Rep. Prog. Phys. 85 056201 https://arxiv.org/pdf/2104.07634.pdf

[13] E. Aprile, XENONnT Analysis: Signal Reconstruction, Calibration and Event Selection, 2024, https://arxiv.org/pdf/2409.08778

[14] Thomas J. Weiler, On the likely dominance of WIMP annihilation to fermion pair+W/Z (and implication for indirect detection), 2012, https://doi.org/10.48550/arXiv.1301.0021

[15] T. Ward, Electroweak Mixing and the Generation of Massive Gauge Bosons, in "Beyond the Desert 2002", edited by H. V. Klapdor-Kleingrothhaus, IOP publishing, Bristol and Philadelphia, 2003. Page 171

[16] National Nuclear Data Center, ENDF/B-VII.1 (2011), https://www.nndc.bnl.gov/endf/

[17] T. Enqvist et al., Measurements of muon flux in the Pyhäsalmi underground laboratory, Nucl. Instrum. Meth. A554 (2005) 286, https://doi.org/10.1016/j.nima.2005.08.065

[18] Cao, Haichuan (2023). Purdue University Graduate School. Thesis. https://doi.org/10.25394/PGS.22685170.v1

[19] Z. Debicki at al., Helium counters for low neutron flux measurements, Astrophys. Space Sci. Trans., **7**, 511-514, 2011

[20] ZdAJ-HITEC webpage: http://hitecpoland.eu Accessed: 2025-01-22

[21] Z. Debicki et al., Thermal neutrons at Gran Sasso. Nucl. Phys. B Proc. Suppl., 196, 429–432, 2009 ,

[22] Z. Debicki at al., Neutron flux measurements in the Gran Sasso national laboratory and in the Slanic Prahova Salt Mine, Nucl. Instrum. Meth. A, 910, 133-138, 2018

[23] Kinga Polaczek-Grelik at al, Natural background radiation at Lab 2 of Callio Lab, Pyhäsalmi mine in Finland, Nucl. Instrum. Meth. A, 969, 164015, 2020

[24] BSUIN webpage: http://bsuin.eu/ Accessed: 2025-01-22

[25] K. Jędrzejczak et al., Environmentally resilient data collecting system optimized for measuring low density neutron flux, Nuclear Instruments and Methods in Physics Research A 1065 (2024) 169493, https://doi.org/10.1016/j.nima.2024.169493

[26] M. Kasztelan, T. Enqvist, K. Jedrzejczak, J. Joutsenvaara, O. Kotavaara, P. Kuusiniemi, K. Loo, J. Orzechowski, J. Puputti, A. Sobkow, M. Slupecki, J. Szabelski, I. Usoskin, W. Trzaska, T. Ward, High-multiplicity neutron events registered by NEMESIS experiment, PoS ICRC2021 (2021) 497, http://dx.doi.org/10.22323/1.395.0497

[27] Z. Dębicki et al., Measurements and interpretation of registration of large number of neutrons generated in lead: the role of particle cascades, Astrophys. Space Sci. Trans., 7, 101–104, 2011

[28] P. Kuusiniemi, et al., Performance of tracking stations of the underground cosmic- ray detector array EMMA, Astropart. Phys. 102 (2018) 67–76, http://dx.doi.org/10.1016/j.astropartphys.2018.05.001

[29] W.H. Trzaska et al., Possibilities for underground physics in the Pyhasalmi mine, in: 13th Conference on the Intersections of Particle and Nuclear Physics, 2018, https://arxiv.org/pdf/1810.00909

[30] W.H. Trzaska et al., NEMESIS setup for Indirect Detection of WIMPs, Nuclear Inst. and Methods in Physics Research, A 1040 (2022) 167223, https://doi.org/10.1016/j.nima.2022.167223

[31] Haichuan Cao and David Koltick, Dark Matter Induced Neutron Production Search Limits, https://arxiv.org/pdf/2506.22659

[32] C.A.Duba et al., HALO: The helium and lead observatory for supernova neutrinos, Journal of Physics: Conference Series 136 (2008) 042077, https://iopscience.iop.org/article/10.1088/1742-6596/136/4/042077

[33] S Al Kharusi et al., SNEWS 2.0: a next-generation supernova early warning system for multi-messenger astronomy, 2021 New J. Phys. 23 031201, https://doi.org/10.1088/1367-2630/abde33

[34] S. Sekula for the HALO Collaboration, Measurements from HALO, XVIII International Conference on Topics in Astroparticle and Underground Physics (TAUP2023) Contribution, https://indi.to/wwdTx

[35] Ming Fang et al., Boron coated straw-based neutron multiplicity counter for neutron interrogation of TRISO fueled pebbles, Annals of Nuclear Energy 187 (2023) 109794, https://doi.org/10.1016/j.anucene.2023.109794

[36] Jeffrey L. Lacy et al., Recent advances in boron-coated straw detectors and electronics for neutron science instruments, 2025, J. Phys.: Conf. Ser. 3130 012014, https://iopscience.iop.org/article/10.1088/1742-6596/3130/1/012014